\documentclass[11pt]{article}

\RequirePackage{calc}
\RequirePackage[american]{babel}

\usepackage{enumerate}
\RequirePackage{fancyhdr}
\RequirePackage{amsmath}
\usepackage{graphicx}
\usepackage{subfigure}

\usepackage{xr}
\usepackage[]{color}

\definecolor{fgcolor}{rgb}{0.345, 0.345, 0.345}

\definecolor{shadecolor}{rgb}{.97, .97, .97}
\definecolor{messagecolor}{rgb}{0, 0, 0}
\definecolor{warningcolor}{rgb}{1, 0, 1}
\definecolor{errorcolor}{rgb}{1, 0, 0}

\usepackage{alltt}

\usepackage{graphicx}
\usepackage{color}
\usepackage{array}
\usepackage{multirow}
\usepackage{mathtools}

\usepackage{float}
\usepackage{bbm}
\usepackage{amsmath}
\usepackage{amssymb}
\usepackage{ulem}
\usepackage{subfig}
\usepackage{wrapfig}
\usepackage{url}

\usepackage{natbib}
\DeclareMathOperator*{\argmin}{argmin}

\newcommand{\Ell}{\mathcal{L}}
\newcommand{\mb}{\mathbf}

\newcommand{\E}{\text{E}}
\newcommand{\V}{\text{Var}}
\newcommand{\cov}{\text{Cov}}
\newcommand{\indfun}[1]{\ensuremath{\mb{1}_{\{#1\}}}}

\newtheorem{theorem}{Theorem}[section]

\newcommand{\qed}{\nobreak \ifvmode \relax \else
      \ifdim\lastskip<1.5em \hskip-\lastskip
      \hskip1.5em plus0em minus0.5em \fi \nobreak
      \vrule height0.75em width0.5em depth0.25em\fi}

\newcommand{\vmdel}[1]{}
\newcommand{\vmadd}[1]{#1}

\begin{document}

\captionsetup[subfigure]{labelformat=empty}

\begin{center}
\Large{Statistical inference in partially observed branching processes with application to 
cell lineage tracking of \textit{in vivo} hematopoiesis}\\\ \\

\normalsize
Jason Xu$^1$, Samson Koelle$^2$, Peter Guttorp$^2$, Chuanfeng Wu$^3$, Cynthia E. Dunbar$^3$,  Janis L. Abkowitz$^{4}$, Vladimir N. Minin$^{5,*}$

\small
$^1$Department of Biomathematics, University of California, Los Angeles\\
$^2$Department of Statistics, University of Washington\\
$^3$National Heart, Lung, and Blood Institute, National Institutes of Health \\
$^4$Department of Medicine, Division of Hematology, University of Washington  \\
$^5$Department of Statistics, University of California, Irvine\\
$^*$Corresponding author; \url{vminin@uci.edu}

\normalsize
\end{center}

\begin{abstract} \noindent 
Single-cell lineage tracking strategies enabled by recent experimental technologies have produced significant insights into cell fate decisions, but lack the quantitative framework necessary for rigorous statistical analysis of mechanistic models describing cell division and differentiation.
In this paper, we develop such a framework with corresponding moment-based parameter estimation techniques for continuous-time, multi-type branching processes. Such processes provide a probabilistic model of how cells divide and differentiate, and we apply our method to study \textit{hematopoiesis}, the \vmdel{complex} mechanism of blood cell production. 
We derive closed-form expressions for higher moments in a general class of such models. 
These analytical results allow us to efficiently estimate parameters of much richer statistical models of hematopoiesis than those used in  previous statistical studies.
To our knowledge, the method provides the first rate inference procedure for fitting such models to time series data generated from cellular barcoding experiments. 
After validating the methodology in simulation studies, we apply our estimator to hematopoietic lineage tracking data from rhesus macaques. 
Our analysis provides a more complete understanding of cell fate decisions during hematopoiesis in non-human primates, which may be more relevant to human biology and clinical strategies than previous findings from murine studies.
\vmadd{For example, in addition to previously estimated hematopoietic stem cell self-renewal rate, we are able to estimate fate decision probabilities and to compare structurally distinct models of hematopoiesis using cross validation.
These estimates of fate decision probabilities and our model selection results should help biologists compare competing hypotheses about how progenitor cells differentiate.}
The methodology is transferrable to a large class of stochastic compartmental models and multi-type branching models, commonly used in studies of cancer progression, epidemiology, and many other fields. 
\end{abstract}

\section{Introduction}\label{sec:intro}
This paper develops inferential tools for a class of hidden stochastic population processes. In particular, we present a correlation-based $M$-estimator for rate inference in multi-type branching process models of \textit{hematopoiesis} \vmadd{---
a mechanism during which self-renewing hematopoietic stem cells (HSCs) specialize, or differentiate, to produce mature blood cells}. 
Understanding the details of this process is a fundamental problem in systems biology, and progress \vmadd{in uncovering these details} will also help shed light on other areas of basic biology. For example,
further advances in hematopoiesis research will yield insights into mechanisms of cellular interactions,
cell lineage programming, and characterization of cellular phenotypes during cell differentiation \citep{orkin2008}.
Moreover, understanding hematopoiesis is clinically important: all
blood cell diseases, including leukemias, myeloproliferative disorders and myelodysplasia are caused by malfunctions in some part of the hematopoiesis process, and hematopoietic stem cell 
 transplantation has become a mainstay for gene therapy and cancer treatments \citep{Whichard2010}.
\par
\vmadd{An HSC can give rise to any mature blood cell. In order to generate new mature blood cells (e.g., granulocytes, monocytes, T, B, and natural killer (NK) cells) an HSC first becomes a multipotent progenitor cell. This cell then further differentiates into progenitors with more limited potential. An HSC can also divide or self-renew, giving rise to two daughter HSCs. Cells make fate decisions by a carefully orchestrated change in gene expression, but the details of these decision making processes are still not fully understood \citep{Laslo2008, Whichard2010}. 
Mathematically, hematopoiesis can be represented \vmdel{mechanistically} as a \textit{stochastic compartmental model} in which cells are assumed to self replicate and differentiate according to a Markov branching process \citep{Becker1963,Siminovitch1963,kimmel2002}.}
\par
\vmadd{Although this mathematical representation of hematopoiesis is more than fifty years old \citep{Till1964}, fitting branching process models to experimental data remains highly nontrivial. 
The main difficulty stems from the fact that estimating parameters of a partially observed stochastic process usually leads to intractable computational algorithms. 
One way to avoid this intractability is to base inference on deterministic models of hematopoiesis, as has been done by \citet{colijn2005} and \citet{marciniak2009}, for example.
However, deterministic compartmental models are not suitable when cells counts are low in some of the compartments \citep{kimmel2014}, which is frequently the case in many experimental protocols (e.g., bone marrow transplantation followed by blood cell reconstitution). 
Although working within the stochastic modeling framework is challenging,
researchers were able to fit a two-compartmental stochastic model to X-chromosome inactivation marker data \citep{Abkowitz1990,golinelli2006, fong2009, Catlin2011}. 
Such studies have produced important insights, but this simple two-compartmental model} cannot distinguish between stages of differentiation beyond the HSC, and results obtained from analyzing this model have not resolved long standing questions about patterns and sizes of cell lineages descended from individual HSC cells. It should be noted that even these simplified models capturing the clonal dynamics descended from an HSC have posed significant statistical and computational challenges. 
\par
Recently emergent experimental techniques now allow researchers to track the dynamics of cell lineages descended from distinct ancestral progenitor or HSC cells. Collecting such high resolution data is made possible by lentiviral genetic 
barcoding coupled with modern high-throughput sequencing technologies \citep{Gerrits2010, Lu2011, wu2014}. Each cell descended from an original barcoded population inherits the unique identifier of its ancestor. The data thus enable us to distinguish individual lineages, and comprise independent and identically distributed time series. This \vmdel{is a} marked departure from previous \vmadd{batch} experimental data \vmdel{observing} \vmadd{, in which observations were coming from} the \vmdel{batch}  population \vmadd{of cells} descended from a mixture of indistinguishable cells, \vmdel{and} potentially allows for investigation of much more realistic models of hematopoiesis. 
Importantly, the ability to analyze individual lineage trajectories can be very useful in characterizing patterns of cell differentiation, shedding light on the larger tree structure of the  differentiation process. 

While these \vmadd{barcoding} data are certainly more informative than those from previous experiments, statistical methods capable of analyzing such data are only beginning to emerge. 
\cite{Perie2014} model genetic barcoding data in a murine study collected at the end of the mice's lifespans, but do not account for the longitudinal aspect of the data. 
\vmadd{They also do not fully take advantage of the information in the read count data, instead working with binary indicators of barcode presence.} 
\citet{goyal2015} present a neutral steady-state model of long term hematopoiesis applied to vector site integration data, but cannot infer crucial process parameters such as the rate of stem cell self-renewal. 
\citet{Biasco2016} manage to estimate cell differentiation rates from blood lineage tracking data, but resort to diffusion approximation and ignore all variation arising from experimental design in their analysis.
\par
\cite{wu2014} provide a preliminary analysis of their cellular barcoding data that reveals important scientific insights \citep{koelle2017}, but lacks the ability to perform statistical tasks such as parameter estimation and model selection. This paper attempts to fill this methodological gap, developing new statistical techniques for studying the barcoded hematopoietic cell lineages from the rhesus macaque data.
We propose a fully generative stochastic model and efficient method of parameter estimation that enables much richer hematopoietic structures to be statistically analyzed than previously possible, allowing models that consider HSC, progenitor, and mature cell stages. The following section describes the model and experimental design producing the dataset we will analyze. Next, we motivate our approach by statistically formulating our inferential goal and deriving the necessary mathematical quantities in Section \ref{sec:methods}. We then thoroughly validate these methods via several simulation studies, \vmdel{and} fit the models to the rhesus macaque barcoding data\vmadd{, and compare the fitted models via cross validation}. Finally, we close with a discussion of these results, their implications, and avenues for future work.

\section{Methods}\label{sec:data}
\subsection{Data and Experimental Setup}
\vmadd{During hematopoiesis, self-renewing hematopoietic stem cells specialize or \textit{differentiate} via a series of intermediate progenitor cell stages to produce mature blood cells \citep{Weissman2000}.  
A challenge in studying this system \textit{in vivo} is that only the mature cells are observable, as they can be sampled from the blood. 
We will model hematopoiesis as a continuous-time stochastic process \vmadd{whose state $\mb{X}(t)$ is a vector of cell counts of different types (e.g., HSCs, progenitors, T, B cells).
We will provide mathematical formulation of the stochastic process} after a complete description of the dataset $\mb{Y}$. 
In contrast to previous studies, the single cell lineage tracking dataset we will analyze opens the possibility of inferring intermediate progenitor behavior. 
We briefly describe the cellular barcoding experiment that makes this possible.}

\citet{wu2014} extract HSC and progenitor cells from the marrow of a rhesus macaque, and use lentiviral vectors to insert unique DNA sequences \vmadd{into the cells} that will each act as an identifying ``barcode." After the extracted cells are labeled in this way, the macaque is \vmdel{then} irradiated so that its residual blood cells are depleted. Next, the labeled cells are transplanted back into the marrow of the animal; reconstitution of its entire blood system is supported from this initial labeled population of extracted cells. All cells descended from a marked cell---its \textit{lineage}--- inherit its unique barcode ID; 
we remark that what we call a lineage is often referred to as a clone in the hematopoiesis literature. 
We assume that \vmadd{barcoded lineages act independently from each other} and \vmadd{that} each barcoded lineage \vmadd{$p \in \{1,\dots,N\}$} is a realization $\mb{X}^p(t)$ of our stochastic process model of hematopoiesis.
\begin{figure}[t]
\centering
\includegraphics[width = .7\paperwidth]{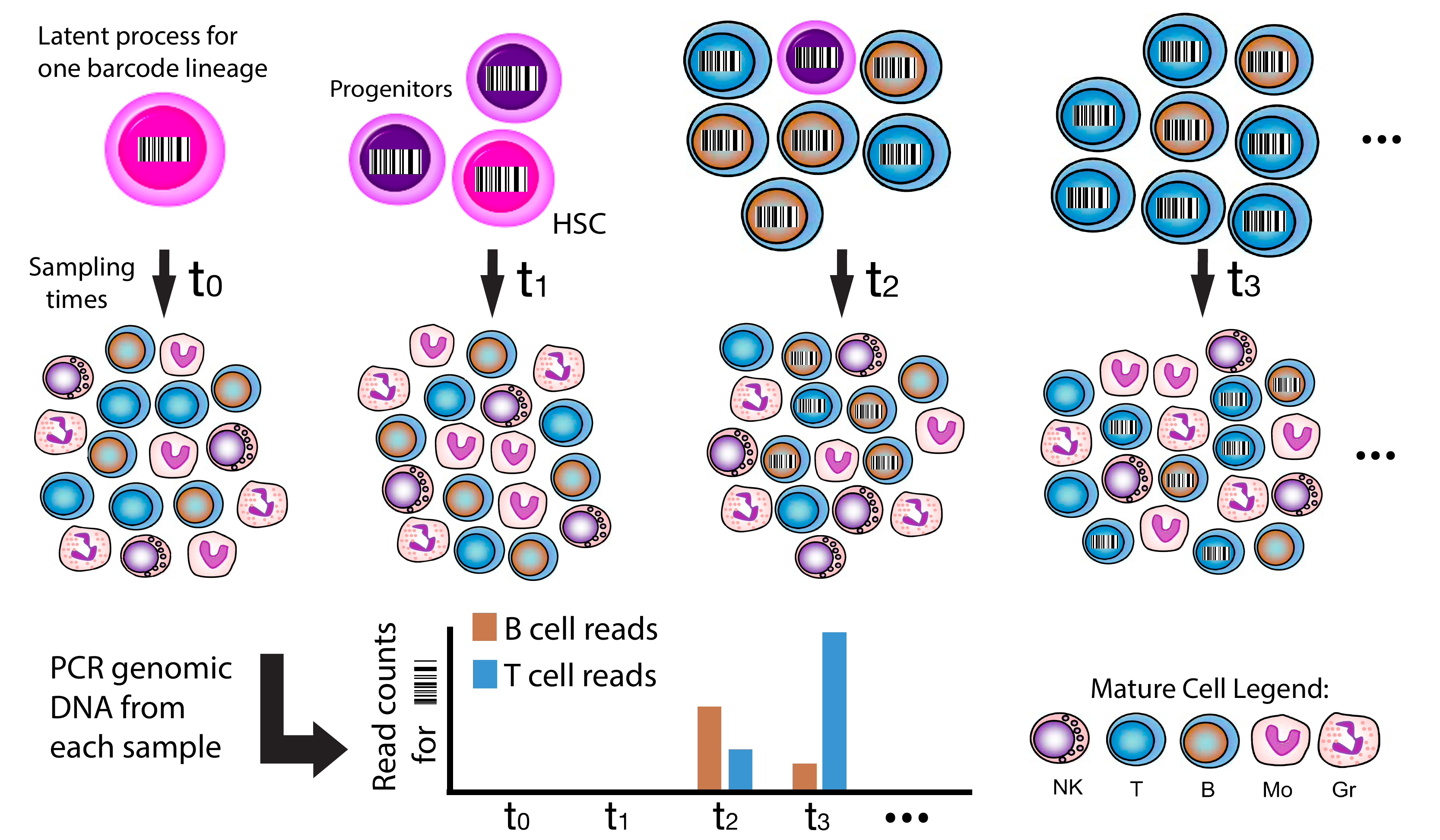}
\caption[Illustration of the experimental protocol generating lineage barcoding data]{ Illustration of experimental protocol for one single fixed barcode ID. The top panel represents the latent process starting with a single HSC (pink) at several snapshots in time $t_0, \ldots, t_3$. The second panel illustrates blood samples. Note that the barcode only becomes present in the blood when mature cells, which first appear by time $t_2$ in this example, are sampled in blood; the HSCs and early progenitors (purple) reside in the marrow and thus are unobservable. Read counts corresponding to the given barcode after PCR and sequencing reflect the number of cells sharing that barcode in the sample, which in turn reflect the barcoded population in the latent process.} 
\label{fig:illustration}
\end{figure}

Hematopoietic reconstitution is monitored indirectly over time at discrete observation times $t_j$. 
At each $t_j$, the experimental protocol consists of sampling blood from the monkey and separating the sample by cell type, followed by retrieving the barcodes via DNA sequencing from each sorted population. 
Specifically, the blood sample is sorted into five mature cell categories: monocyte (Mono), granulocyte (Gr), T, B, and natural killer (NK) cell types. These type-sorted samples will be denoted $\widetilde{\mb{y}}_m(t_j)$ for each mature cell type $m$, and are of fixed size $b_m$ at all observation times. That is, each entry $\widetilde{y}^p_m(t_j)$ is the number of \vmadd{type $m$} cells with barcode $p$ present in the sample, and $\sum_p \widetilde{y}^p_m(t_j) = b_m$ at every $t_j$. The random number of barcodes present in the samples is proportional to their prevalence in the total population of labeled type $m$ cells in the population, denoted $B_m(t_j) = \sum_p X^p_m(t_j)$, where $X^p_m(t_j)$ denotes \vmadd{the true blood count} of type $m$ cells \vmadd{from lineage $p$} at time $t_j$. 
Therefore, the distribution of sampled cells can be modeled by a multivariate hypergeometric distribution
\begin{equation}
\widetilde{\mb{y}}_m(t) \mid \mathbf{X}(t) \sim \text{mvhypergeom}(B_m(t), \mb{X}_m(t), b_m).
\end{equation}
\vmadd{Put another way, 
$\text{Pr}(\widetilde{y}_m^p(t) = z)$ is the probability of drawing 
 $z$ balls of color $p$ out of an urn containing $B_m$ total balls, $X_m^p(t)$ of which are of color $p$, in a sample of size $b_m$}. 
 In this setting, each color corresponds to a barcode ID; the distributional choice is driven by its close mechanistic resemblance to the experimental sampling itself. 
Recall that the sample sizes $b_m$ are fixed and known from the experimental protocol. While the latent processes are unknown, the values of their sum $B_m(t_j)$ are observed: the total circulating blood cell (CBC) counts are \vmdel{measured} \vmadd{recorded} at each sampling time. We do not consider potential measurement error in the CBC data, and therefore do not model $B_m(t_j)$ as random variables throughout, instead treating $B_m(t_j)$ as external known constants on which we condition $\mb{X}^p(t)$. 

Next, individual barcodes must be retrieved or \textit{read} via sequencing. \vmadd{DNA is extracted from each of the sorted samples, and polymerase chain reaction (PCR) is performed to generate many copies of the DNA segments. This step aids barcode retrieval by increasing detectability of DNA segments present in the sample during sequencing. It is commonly assumed that PCR amplification  preserves the proportion of barcodes present. We disregard experimental noise that may cause negligible departures from this standard assumption in order to avoid modeling PCR itself as an additional stochastic process.} The read count $y_m^p(t)  = d_m(t_j) \times \widetilde{y}^p_m(t_j)$ is obtained by sequencing this amplified PCR product; here $d_m(t)$ is an unknown constant representing the linear effect of PCR amplification.  Thus, at each observation time $t_j$, the experiment yields a count $y_m^p(t_j)$ denoting the number of times barcode ID $p$ was read after sequencing the type $m$ cell sample. 
An illustration summarizing the process for one lineage is provided in Figure \ref{fig:illustration}.

Our assumption that PCR amplification is linear may be less suitable for barcoded populations with low counts, as any noise we have chosen not to model from this process may have a larger relative effect.
We therefore further filter the data similarly to \citep{wu2014} to include only barcode IDs exceeding $1000$ reads. 
\vmadd{Altogether, our observed dataset consists of over $110$ million read counts across $N= 9635$ unique barcode IDs, obtained at \vmadd{irregularly spaced times}  over a total period of $t_J=30$ months. 
This collection of read counts can be viewed as a three-dimensional array, where the first array index $m$ corresponds to mature cell type $m$.
\vmadd{Fixing this index results in a $N \times J$ matrix $\mb{Y}_m = ( \mb{y}_m(t_1), \mb{y}_m(t_2), \ldots, \mb{y}_m(t_j))$}. The second array index, columns of each such matrix described above, correspond to observation (sampling) times $\mb{t} = (t_1, \ldots, t_J)$. The third array dimension indexes barcodes:  $\mb{y}^p_m$, the $p$th row of $\mb{Y}_m$, encodes the read count time series corresponding to a unique barcode ID $p \in \{1, \ldots, N\}$ among the population of type $m$ mature blood cells.} 

\subsection{Multi-Type Branching Model of Latent Process}\label{sec:branchmodel}
The data $\mb{Y}$ form a \textit{partial observation} of a collection of $p$ IID continuous-time latent processes, each evolving according to the stochastic model $\mb{X}(t)$. We now provide a biological description of the underlying hematopoietic process we wish to model by $\mb{X}(t)$, followed by mathematical details of our proposed class of branching process models. 

\vmadd{Hematopoiesis begins with bone marrow residing HSCs, which have the capacity to self-renew (give rise to another HSC) or differentiate into more specialized progenitor cells. Biologists have not reached a consensus about how many types of progenitors exist in this intermediate stage, but agree that intermediate progenitor cells lose the ability to proliferate, and each progenitor type can produce one or several types of mature blood cells before exhausting its own lifespan (cell death). These mature blood cells exist at the last phase of development, are found mainly in the bloodstream and do not give birth to any further cells. 
Based on this biological understanding of hematopoiesis, a multi-type branching process taking values over a discrete state space of cell counts in continuous time provides a natural modeling choice.
Canonical differentiation trees that have been posited in the scientific literature follow such a structure, and such stochastic models have established their place in the statistical hematopoiesis literature through several studies \citep{kimmel2002, Catlin2011}. }

\begin{figure}[htbp]
\centering
\subfigure[]{
    \includegraphics[scale=.435]{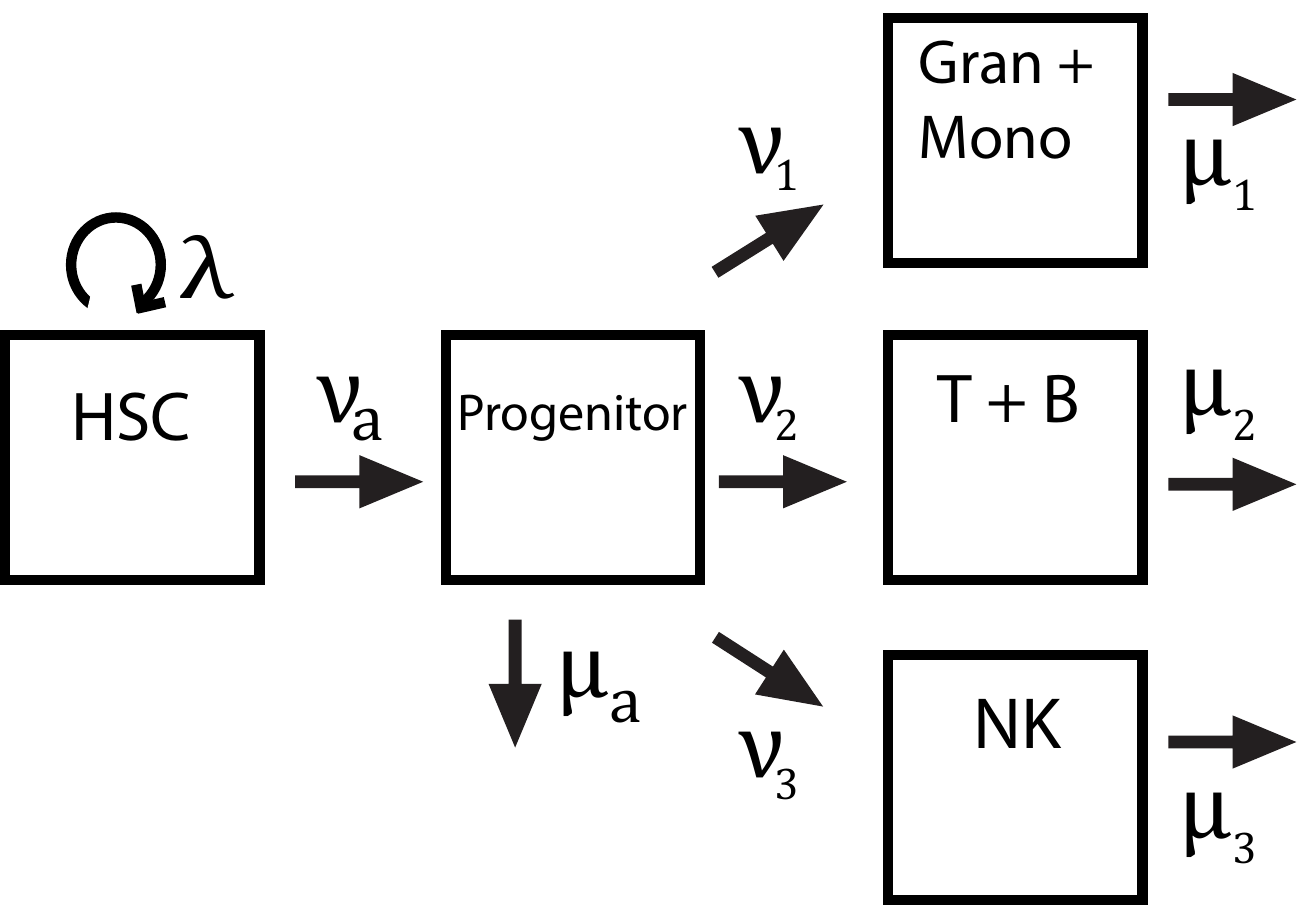}
} 
\subfigure[]{
    \hspace{2pt}
    \includegraphics[scale=.415]{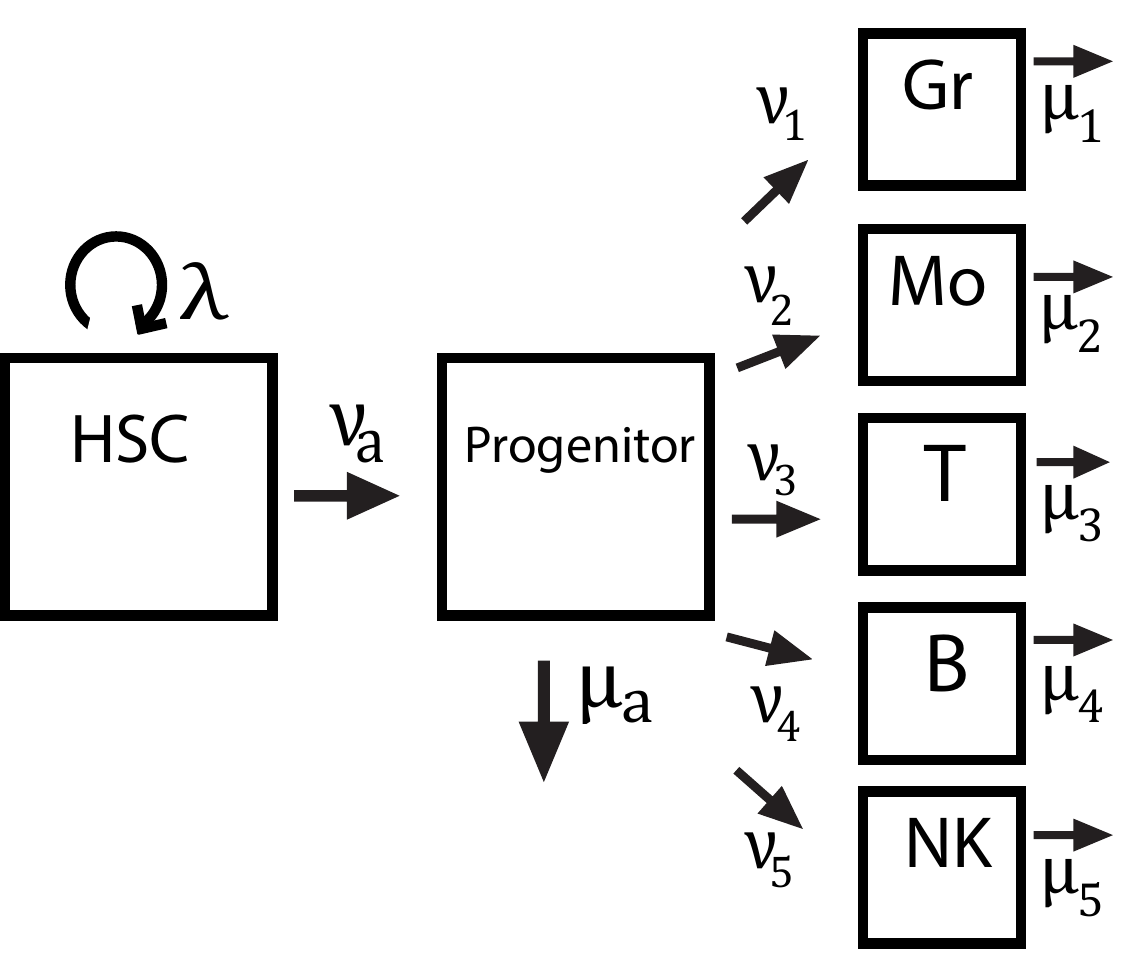}
} 
\subfigure[]{
    \hspace{2pt}
    \includegraphics[scale=.415]{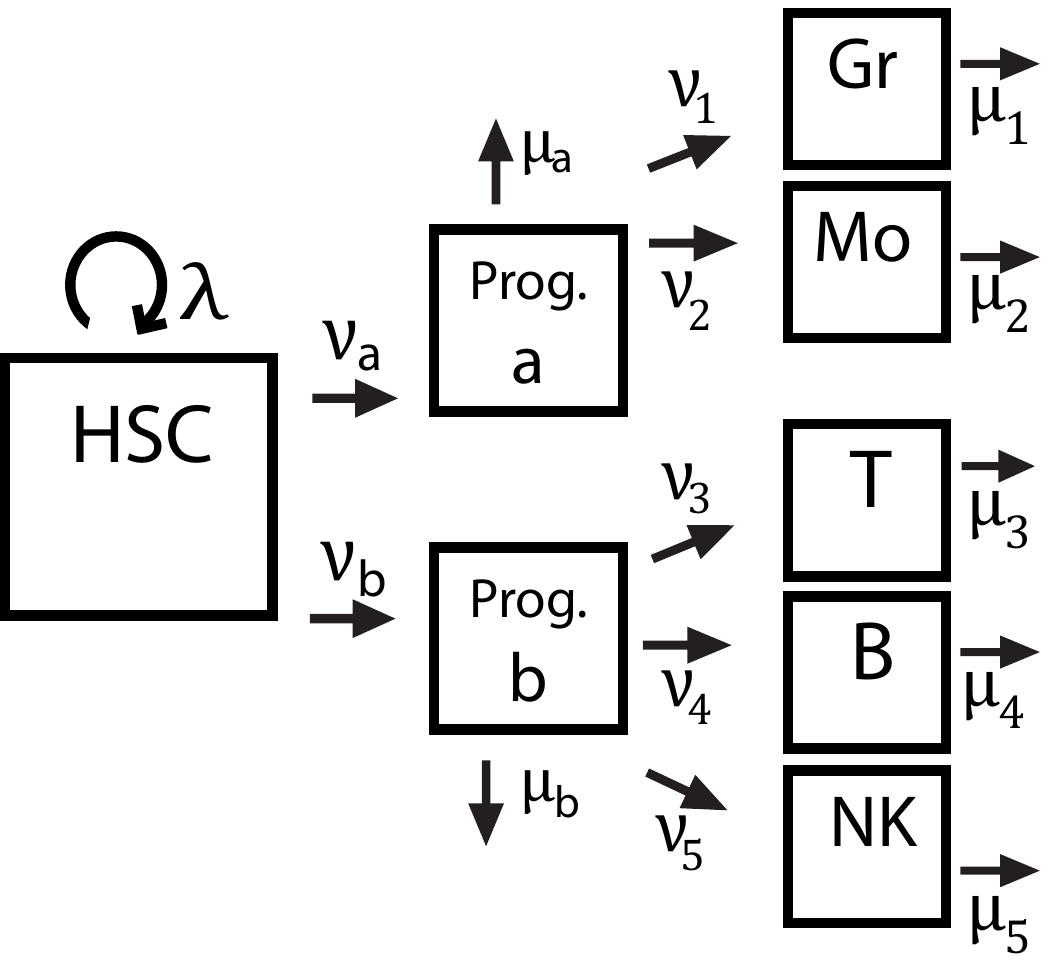}
} 
\centering
\subfigure[]{
    \includegraphics[scale=.455]{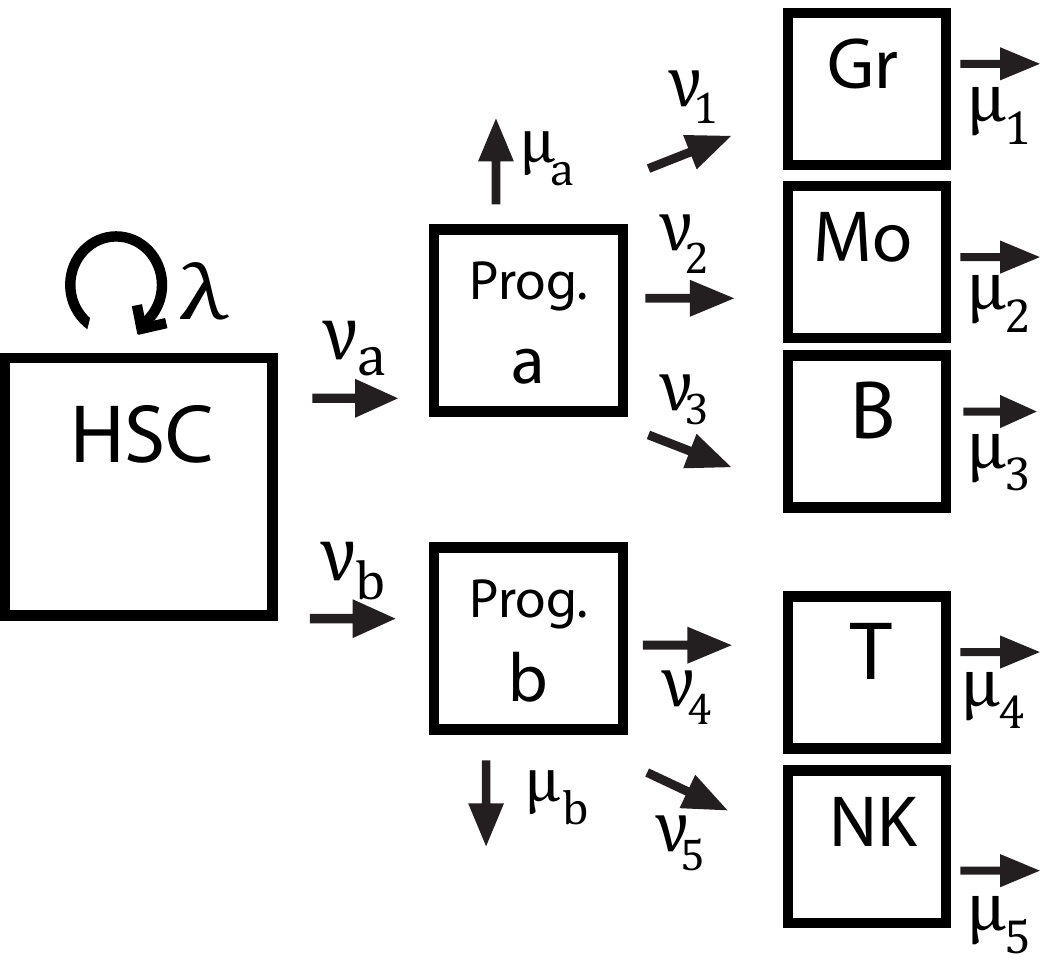}
} 
\subfigure[]{
    \hspace{3pt}
    \includegraphics[scale=.455]{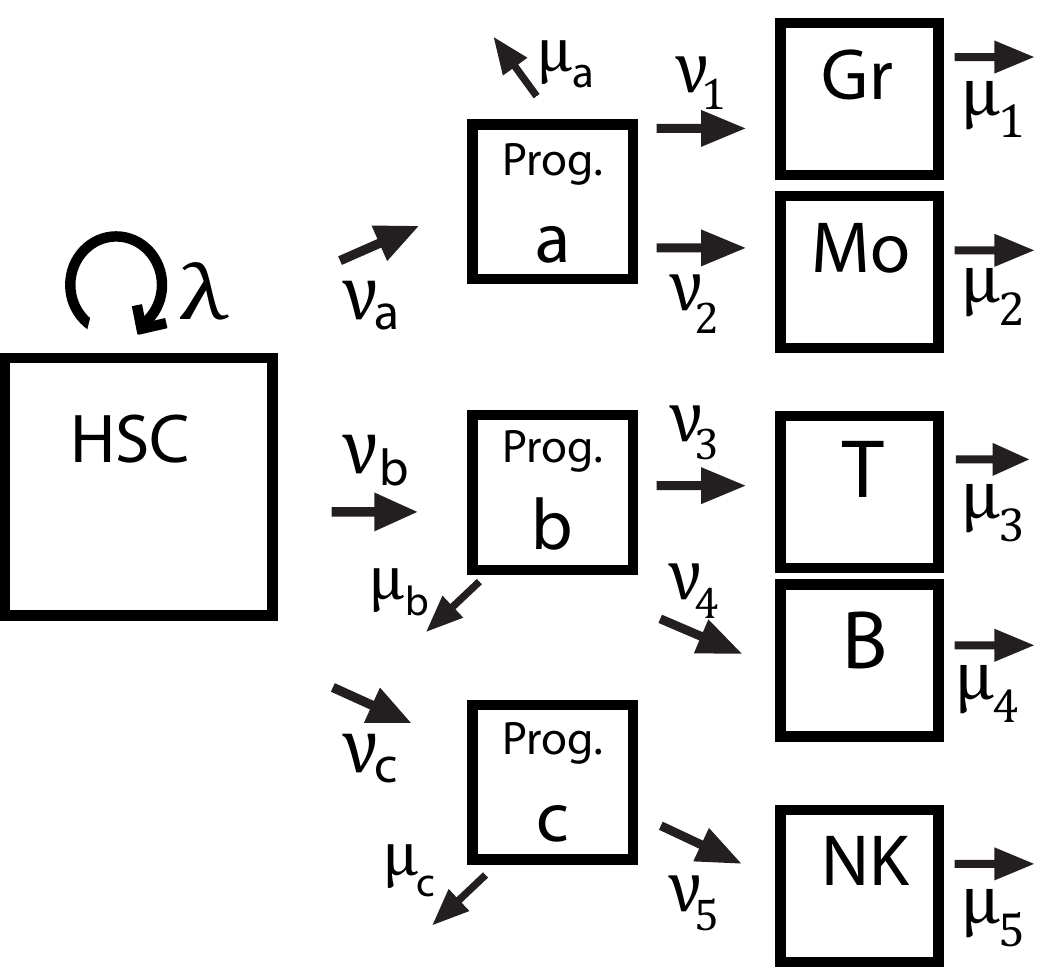}
} 
\subfigure[]{
    \hspace{3pt}
    \includegraphics[scale=.455]{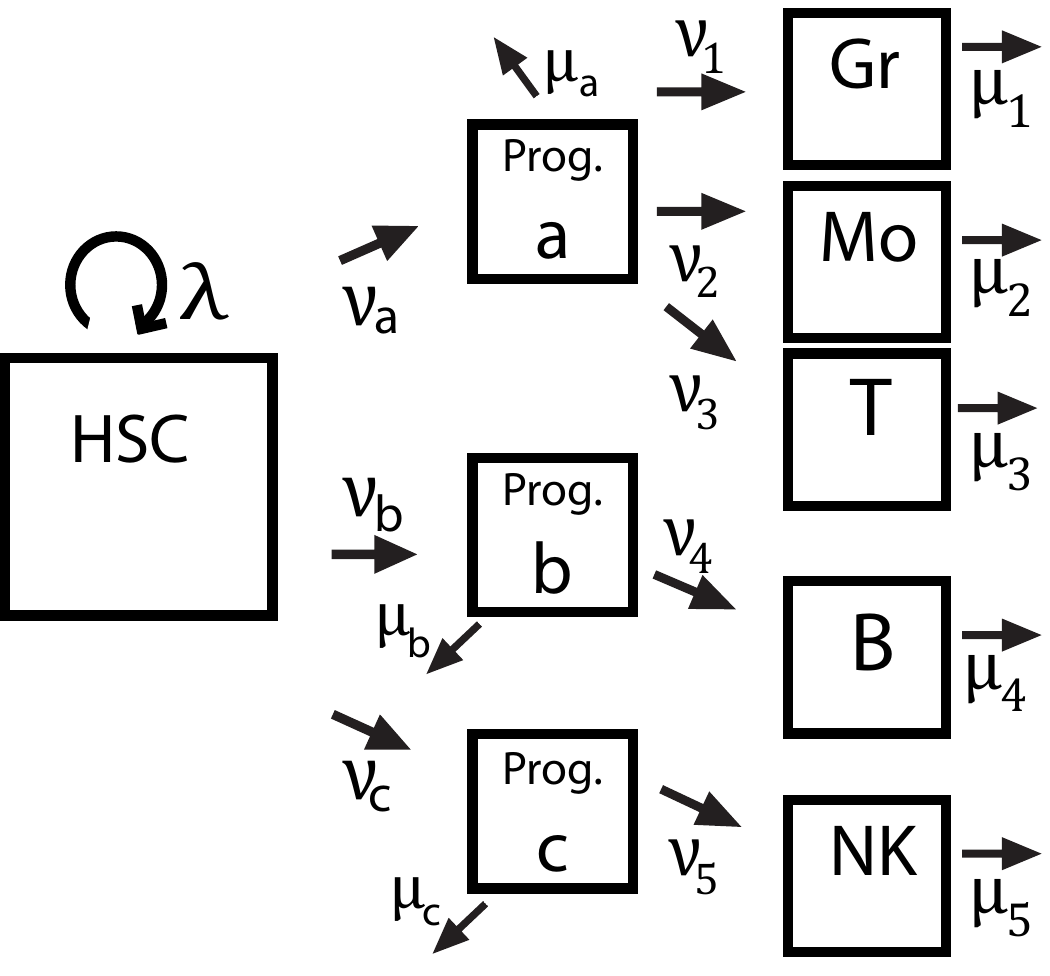}
} 
\caption[Model diagrams in the class of branching processes we consider]{Differentiation trees to be considered in simulation study and real data analysis. In the first two models, mature cells are descended from one common \textit{multipotent} progenitor: (a) groups similar mature cells (i.e. T and B cells are not distinguishable), leading to a model with three mature cell types, and (b) models each observed mature cell type separately. Note that previous statistical studies by \cite{catlin2001, golinelli2006, fong2009} model only two compartments (types), containing HSC and ``other" cells. Models (c)---(f) include several biologically plausible topologies featuring two or three \textit{oligopotent} progenitors, each specializing to produce only particular mature cells. } 
\label{fig:allmodels}
\end{figure}

A continuous-time branching process is a Markov jump process in which a collection of independently acting particles (cells) can reproduce and die according to a probability distribution. Each cell type has a distinct mean lifespan and reproductive probabilities, and can give rise to cells of its own type as well as other types at its time of death. 
Our branching process models consist of an HSC stage, progenitor stage, and mature cell stage, and allow \vmadd{for} an arbitrary number of progenitor and mature cell types to be specified. We use alphabetic subscripts $a, b \ldots \in \mathcal{A}$ to denote progenitors, with mature cell types indexed numerically by $m=1, 2, \ldots \mathcal{M}$. The subscript $0$ indicates quantities relating to HSCs.  In our models, HSCs self-renew with rate $\lambda$, or become type $a$ progenitor cells with differentiation rates $\nu_a$. 
Progenitor cells exhaust their lifespan with rates $\mu_a$, and produce type $m$ mature blood cells with rates $\nu_m$. 
Each mature cell type $m$ is descended from only one progenitor type, so that its corresponding production rate $\nu_m$ is unique and well-defined. Finally, these mature cells exhaust with rates $\mu_m$. Figure~\ref{fig:allmodels} depicts several example structures contained in this class. 
In a given branching model, let $C = 1 + |\mathcal{A}| + \mathcal{M}$ be the total number of cell types.
\vmadd{
The process state is a length $C$ random vector $\mb{X}(t) = (X_0(t), X_a(t), \ldots, X_{|\mathcal{A}|}(t),  X_1(t), \ldots, X_{\mathcal{M}}(t) )$ taking values in the countably infinite state space $\Omega = \mathbb{N}^C$, whose components represent sizes of the cell populations \vmdel{of the corresponding subscript} at time $t \geq 0$. Recall the read count data $\mb{Y}$ are obtained by sequencing blood samples from the mature populations $X_1(t), \ldots, X_{\mathcal{M}}(t)$; the early stage populations $X_0(t), X_a(t), \ldots, X_{|\mathcal{A}|}(t)$ are entirely unobserved.}

The behavior of $\mb{X}(t)$ is then defined by specifying a set of length $C$ \textit{instantaneous rate vectors}. 
To introduce the remaining notation, we focus on the simplest model displayed in Figure~\ref{fig:allmodels}(a) with $C=5$ total cell types for concreteness. Model ~\ref{fig:allmodels}(a) features one progenitor type and three mature cell types.  
The instantaneous rate $\alpha_0(n_0, n_a, n_1, n_2, n_3)$ \citep{dorman2004,lange2010} contains the rate of an event occurring in which an HSC cell produces $n_0$ HSCs, $n_a$ progenitors, and $n_m$ of each type $m$ mature cells. 
These rate vectors are analogous \vmdel{beginning with} \vmadd{for} other \vmadd{parent} cell types: for instance, $\alpha_a(n_0, n_a, n_1, n_2, n_3)$ denotes the same rates of production from one type $a$ progenitor cell rather than an HSC cell. The offspring  descended from each cell subsequently behave according to the same set of rate vectors, which do not change with $t$ --- the process is \textit{time-homogeneous}.

The assumption that cells act independently implies that the process rates are \textit{linear}: overall event rates at the population level are multiplicative in the number of cells.
Together, these assumptions imply that the lifespan of each HSC follows an exponential distribution with parameter $- \beta_0 := \sum_{(n_0, n_a, n_1, n_2, n_3) \neq (1,0,0,0,0)} \alpha_0(n_0, n_a, n_1, n_2, n_3)$. After this exponential waiting time, the probability that the cell is replaced by $(n_0, n_a, n_1, n_2, n_3)$ cells of each respective type is given by normalizing the corresponding rate: $\alpha_0(n_0, n_a, n_1, n_2, n_3)/\beta_0$. The same holds analogously for all cells after replacing subscripts appropriately, and therefore $\mb{X}(t)$ evolves over time as a continuous-time Markov chain (CTMC) \citep[Chapter 3]{guttorp1995}.

As an example, we see from Figure \ref{fig:allmodels} that model \ref{fig:allmodels}(a) is characterized by the parameters $\boldsymbol{\theta} = (\lambda, \nu_a, \mu_a, \nu_1, \nu_2, \nu_3, \mu_1, \mu_2, \mu_3)$. Specifying such a process as a CTMC classically using the rate matrix (infinitesimal generator) is mathematically unwieldy --- this is an infinite matrix with no simplifying structure for these models. 
However, the process can be compactly specified using the following instantaneous rate vectors of a branching process:
$$ \alpha_0(2,0,0,0,0) = \lambda, \quad \alpha_0(0,1,0,0,0) = \nu_a, \quad \alpha_0(1,0,0,0,0) = -(\lambda + \nu_a), \quad \alpha_a(0,0,0,0,0) = \mu_a, $$
$$\alpha_a(0,1,1,0,0) = \nu_1, \quad \alpha_a = (0,1,0,1,0) = \nu_2, \quad \alpha_a(0,1,0,0,1) = \nu_3, \quad \alpha_a(0,1,0,0,0) = -(\mu_a + \nu_1 + \nu_2 + \nu_3),$$
$$\alpha_1(0,0,0,0,0) = \mu_{1}, \quad \alpha_1(0,0,1,0,0) = -\mu_{1}, \quad \alpha_2(0,0,0,0,0) = \mu_2, $$
$$\alpha_2(0,0,0,1,0) = -\mu_2, \quad \alpha_3(0,0,0,0,0) = \mu_3, \quad \alpha_3(0,0,0,0,1) = -\mu_3,$$ with all other rates zero. An equivalent representation in terms of chemical kinetic rate notation is provided in the Appendix A.1.
Given that the instantaneous branching rates of each model can be specified this way from \vmadd{the parameter vector} $\boldsymbol\theta$, methods of inference in the next section target $(\boldsymbol\theta, \boldsymbol\pi)$, where $\boldsymbol\pi$ is an initial distribution vector $\boldsymbol\pi = (\pi_0, \pi_a, \pi_b, \ldots)$. 
The components $\pi_a$ represents the probability that a lineage is originally descended from a \vmadd{transplanted} progenitor rather than from a \vmadd{transplanted} HSC: this is unknown since the initial barcoding is applied to a heterogeneous transplanted cell population containing HSCs and early progenitors. 

\subsection{Parameter Estimation Procedure}\label{sec:methods}
\par
\vmadd{We estimate model parameters using} the generalized method of moments, 
a computationally simpler alternative to maximum likelihood estimation that yields consistent estimators. Perhaps more appealing than their simplicity, moment-based methods feature more robustness to model misspecification than techniques relying on a completely prescribed likelihood \citep{wakefield2013}. The choice is well-motivated when a large number of samples is available, as is the case for our dataset consisting of thousands of IID barcoded lineages.
The method relies on deriving equations relating a set of population moments to the target model parameters to be estimated. 
Next, the discrepancy between the population and sample moments is minimized to estimate parameters of interest. Our estimator seeks to match pairwise empirical read count correlations across barcodes with their corresponding model-based population correlations. We derive explicit analytic forms for the first and second moments of our general class of branching models for hematopoiesis, allowing for the computation of marginal correlations between any pair of mature types. 
The advantage of working with correlations in the data is twofold: first, the observed correlation profiles between types are more time-varying and thus more informative than the mean and variance curves of read counts. Second, the scale invariance of correlations \vmdel{motivates the choice to} \vmadd{allows us to} avoid modeling PCR amplification on top of an already complex model, as the amplification constants $d_m(t)$ cancel out.  
This robustness comes with a caveat --- we may not expect all branching process rates to be identifiable with a scale free approach, instead requiring some parameters be fixed to provide scale information. This will be further discussed in Section \ref{sec:sim}.

With closed form moment expressions, model-based correlations can be computed very efficiently given any parameters $\boldsymbol\theta$, enabling the use of generic optimization methods to minimize a loss function relating the model-based correlations to observed correlations in the read count data. 
The following derivations apply to all models in the class described in Section \ref{sec:branchmodel}, including all those depicted in Figure \ref{fig:allmodels}. 

\subsection{Correlation Loss Function}
To estimate the parameter vector $\boldsymbol\theta$ containing process rates and initial distribution $\boldsymbol{\pi}$, we seek to closely match model-based correlations to the empirical correlations between observed read counts. This is achieved by minimizing the loss function 
\begin{equation}
\Ell(\boldsymbol{\theta}; \mb{Y})  =  
 \sum_{t_j} \sum_m \sum_{n \neq m} \left[ \psi_{mn,j}(\boldsymbol{\theta}; \mb{Y} ) - \hat\psi_{mn,j}( \mb{Y} ) \right]^2,
 \label{eq:momentsObjCorr} 
 \end{equation}
 where $\psi_{mn,j}$ represents model-based correlation between reads of type $m,n$ mature cells at time $t_j$:
 \[ \psi_{mn,j}(\boldsymbol{\theta}; \mb{Y} ) :=   \rho ( Y_m(t_j), Y_n(t_j) ; \boldsymbol{\theta} )= 
 \frac{ \text{Cov}[Y_m(t_j), Y_n(t_j); \boldsymbol{\theta}]}{ \sigma(Y_m(t_j);\boldsymbol{\theta}) \sigma(Y_n(t_j);\boldsymbol{\theta}) }, \]
and $\hat\psi_{mn,j}$ denotes corresponding sample correlations across barcodes $p=1,\ldots, N$ at time $t_j$:
 \[  \hat\psi_{mn,j}(\mb{Y}) :=  \hat\rho(\mb{y}_m(t_j), \mb{y}_n(t_j) ) =  \frac{ \sum_{p=1}^N (y_m^p(t_j) - \overline y_m(t_j) ) (y_n^p(t_j) - \overline y_n(t_j) ) }{  \sqrt{ \sum_{p=1}^N (y_m^p(t_j) - \overline y_m(t_j) )^2} \sqrt{ \sum_{p=1}^N (y_n^p(t_j) - \overline y_n(t_j) )^2} } .
  \]
The problem of estimating hematopoietic rates now translates to seeking 
\[ \hat{\boldsymbol{\theta}}_N = \argmin_{\boldsymbol{\theta}}\Ell(\boldsymbol{\theta}; \mb{Y}) = \argmin_{\boldsymbol{\theta}} \lVert \mb{G}_N(\boldsymbol{\theta}; \mb{Y}) \rVert_2^2 , \qquad \text{where } \mb{G}_N(\boldsymbol{\theta}; \mb{Y}) := \boldsymbol\psi(\boldsymbol{\theta};\mb{Y}) - \hat{\boldsymbol\psi}(\mb{Y}), \]
and $\boldsymbol\psi(\boldsymbol{\theta};\mb{Y}), \hat{\boldsymbol\psi}(\mb{Y})$ are vectors containing all pairwise model-based and empirical correlations at each time point, respectively. 
If $\boldsymbol\theta_0$ are the true data generating parameters, then $\E [\mb{G}_N(\boldsymbol{\theta_0}; \mb{Y})] \rightarrow 0$ as the number of processes $N \rightarrow \infty$. Our method is therefore an \textit{$M$-estimator}, also known as generalized method of moments (GMM) \citep[Chapter 5]{van2000}.
$M$-estimators are consistent under standard assumptions, as stated in the following theorem. Details and regularity conditions are included in the Appendix A-2.  

\begin{theorem} Under regularity conditions A1---A3 (see Appendix A.2), the sequence $ \{ \hat{\boldsymbol\theta}_N \}$ converges in probability to $\boldsymbol\theta_0$, where $\hat{\boldsymbol{\theta}}_N = \argmin_{\boldsymbol{\theta}}\Ell(\boldsymbol{\theta}; \mb{Y})$, $\Ell(\boldsymbol{\theta}; \mb{Y}) =  \lVert \mb{G}_N ( \boldsymbol\theta) \rVert_2^2 $, and $N$ is the number of independent processes or rows in $\mb{Y}$. \end{theorem}

In addition to serving as a useful context for analyzing properties of $\hat{\boldsymbol{\theta}}_N$, it is worth mentioning that the GMM framework provides a natural extension of our loss function estimator by replacing the $\ell^2$ norm $\lVert \cdot \rVert_2$ by a general family of norms $\lVert \cdot \rVert_W$ induced by positive definite weight matrices $\mb{W}$. The estimator is now given by
\[ \hat{\boldsymbol{\theta}}_W = \argmin_{\boldsymbol{\theta}} \lVert  \mb{G}_N(\boldsymbol{\theta}; \mb{Y}) \rVert_W^2
:=  \argmin_{\boldsymbol{\theta}} \mb{G}_N(\boldsymbol{\theta}; \mb{Y}) ^T \, \hat{\mb{W}}\, \mb{G}_N(\boldsymbol{\theta}; \mb{Y}) ; \]
notice minimization of $\Ell(\boldsymbol{\theta}; \mb{Y})$ is the special case of $\hat{\mb{W}} = \mb{I}$. The norm induced by $\mb{W}$ allows different moment equations to have unequal contributions to the objective function, and its estimate $\hat{\mb{W}}$ from the data intuitively assigns less weight to components which have higher variance and thus provide less information.  GMM estimators $\hat{\boldsymbol{\theta}}_W$ enjoy asymptotic normality under additional regularity assumptions which we do not impose here \citep{pakes1989,van2000}, and are furthermore asymptotically efficient under optimal choice of $\hat{\mb{W}}$ \citep{hansen1982}.  While many algorithms exist for estimating the weight matrix $\hat{\mb{W}}$, the task is nontrivial \citep{hansen1996}. 
Because we have a large enough dataset such that finite-sample efficiency is of lesser concern and we do not expect particular time points or correlation pairs to be more informative than others, we opt for the simple case with $\hat{\mb{W}} = \mb{I}$, avoiding the inclusion of many additional entries of the weight matrix as parameters to be estimated.

Having established the data generating model and estimation framework, 
next we derive the second moments of the latent process $\mb{X}(t)$ using branching process techniques. While this enables us to compute model-based correlations of the branching process, we must then relate these quantities to those in the observed process $\mb{Y}$(t): we do so by connecting correlations of $\mb{X}(t)$ and $\mb{Y}(t)$  via laws of iterated expectations and (co)variances. 

\subsection{Moments of the Multi-Type Branching Process}
Here we derive analytic expressions for the first and second moments of the latent branching processes defined in Section \ref{sec:branchmodel}, enabling efficient computation of model-based correlations $\psi_{mn,j}(\boldsymbol\theta, \mb{Y})$ appearing in the loss function. For quantities relating to all cell types, we will use the common index $i = 0, a, b, \ldots,1, \ldots,\mathcal{M}$, and adopt the notation $\mb{e}_i$ to represent the vector of length $C$ (denoting total number of cell types) with a $1$ in the type $i$ component and is $0$ elsewhere. The indicator $\indfun{a \rightarrow m}$ equals $1$ if mature cell type $m$ is descended from progenitor type $a$ in a given model, and $0$ otherwise.

Our approach is similar to the random variable technique introduced by \cite{bailey1964}, but we derive expressions by way of probability generating functions rather than appealing to the cumulants.
We begin by writing the \textit{pseudo-generating functions}, also called progeny generating functions \citep{dorman2004}, defined as
\begin{equation}\label{eq:pseudogen} u_i(\mb{s}) = \sum_{k_0} \sum_{k_a} \cdots \sum_{k_\mathcal{M}} a_i(k_0, \ldots, k_\mathcal{M}) s_0^{k_0} s_a^{k_a} \cdot s_{\mathcal{M}}^{k_\mathcal{M}}, \end{equation}
where $\mb{s}$ is a vector of dummy variables confined to the $[0,1]$ interval. 
For our class of models, these are given by \begin{align*}
 u_0(\mb{s}) &= \lambda s_0^2 + \sum_{a \in \mathcal{A} } \nu_a s_a  - \left(\lambda + \sum_{a \in \mathcal{A}} \nu_a \right) s_0 , \\
 u_a(\mb{s}) &=  \sum_{m=1}^\mathcal{M} \nu_m s_a s_m \indfun{a \rightarrow m} + \mu_a - \left(\mu_a + \sum_{m=1}^\mathcal{M}\nu_m \indfun{a \rightarrow m} \right) s_a   \quad \quad \forall a \in \mathcal{A} , \\
  u_m(\mb{s}) &= u_m(s_m) = \mu_m - \mu_m s_m, \quad \quad \quad \forall m = 1, \ldots, \mathcal{M} .
 \end{align*}
Next, we can write the probability generating function (PGF) of the process, beginning with one HSC cell, via a relation to the pseudo-generating function $u_0$ as follows:
\begin{align}
\phi_0(t; \mb{s}) &= \E\left[ \prod_{i} s_i^{X_i(t)} | \mb{X}(0) = \mb{e}_0 \right] = \sum_{k_0=0}^\infty \cdots \sum_{k_{\mathcal{M}}=0}^\infty \text{Pr}_{\mb{e}_0,(k_0, k_a, \ldots k_{\mathcal{M})}}(t) s_0^{k_0} s_a^{k_a} \cdots s_{\mathcal{M}}^{k_{\mathcal{M}}} \nonumber \\
&= \sum_{k_0=0}^\infty \cdots \sum_{k_\mathcal{M}=0}^\infty \left[ \mathbf{1}_{ \{ k_0=1, k_a = \ldots = k_{\mathcal{M}} =0 \} } + a_0(k_0, \ldots, k_{\mathcal{M}}) t + o(t) \right] s_0^{k_0} s_a^{k_a} \cdots s_{\mathcal{M}}^{k_{\mathcal{M}}} \nonumber \\
&= s_0 + u_0(\mb{s}) t + o(t). \label{eq:pgfHema}
\end{align}
Analogously defining  $\phi_i$ for processes beginning with one type $i$ cell ($i=1, \ldots, C$), Equation \eqref{eq:pgfHema} yields the relation
\[ \frac{\partial}{\partial t} \phi_i(t, \mb{s}) = u_i (\phi_0(t, \mb{s}), \ldots, \phi_\mathcal{M}(t, \mb{s}) ). \]
Only expressions conditioning on one initial cell are required throughout, since each latent process represents cells sharing a unique genetic barcode, which is always descended from a single marked cell.
Now, let $M_{m|i}(t)$ denote the expected number of type $m$ cells at time $t$, given one initial type $i$ cell. From definition of $\phi_i$, we see that we can relate the probability generating functions to these first moments via partial differentiation:
\[ M_{m | i}(t) = \frac{ \partial}{\partial s_m} \phi_i(t, \mb{s}) |_{s_0 = s_a = \ldots = s_\mathcal{M} = 1}  . \]
Similarly, we may further differentiate the PGF to derive second moments used toward variance and covariance calculations. The relationship $U_{mn|i}(t) = \frac{\partial^2 \phi_i}{\partial s_m \partial s_n} \bigg|_{\mb{s} = \mb{1}}$ holds, where 
\[ U_{mn|i}(t) := \E \left[ X_m (X_n - \indfun{m=n} ) | \mb{X}(0) = \mb{e}_i \right] . \] 
These identities via partial differentiation enables us to write a system of differential equations governing the moments. Applying the multivariate chain rule and the Fa{\`a} di Bruno formula,
\begin{equation}\label{eq:genfirst} \frac{\partial}{\partial t} M_{m | i}(t) = \frac{ \partial^2 \phi_i}{\partial t \partial s_m} \bigm|_\mb{s = 1} = \sum_{k} \frac{\partial u_i}{\partial s_k} \frac{ \partial \phi_k}{\partial s_m} \bigm|_\mb{s = 1}, \end{equation}
\begin{equation}\label{eq:gensecond} \frac{\partial}{\partial t} U_{mn|i}(t) = \frac{ \partial^3 \phi_i}{\partial t \partial s_m \partial s_n} \bigm|_\mb{s = 1} = \sum_{j= 1} \left(\frac{\partial u_i}{\partial \phi_j} \frac{ \partial^2 \phi_j}{\partial s_m \partial s_n}\right) + \sum_{j,k=1} \left( \frac{\partial^2 u_i}{\partial \phi_j \partial \phi_k} \frac{\partial \phi_j}{\partial s_m} \frac{\partial \phi_k}{\partial s_n} \right) \bigm|_\mb{s = 1} . \end{equation}
Notice equation \eqref{eq:genfirst} defines a system of ordinary differential equations (ODEs) determining the mean behavior, whose solutions can be plugged in to solve the second system of equations \eqref{eq:gensecond} governing second moments. These systems are subject to the initial conditions $M_{m|i}(0) = \indfun{m=i}$, $U_{mn|i}(0) = 0$.
We introduce the notation $\kappa_{ij} = \frac{\partial u_i}{\partial s_j} \bigm|_\mb{s=1}$ for brevity; as an example, for all $a \in \mathcal{A}, m =1,\ldots, \mathcal{M}$,
\[ \kappa_{00} = \lambda - \sum_{a \in \mathcal{A}} \nu_a, \quad \kappa_{aa} = -\mu_a, \quad \kappa_{mm} = -\mu_m, \quad \kappa_{0a} = \nu_a, \quad \kappa_{am} = \nu_m \indfun{a\rightarrow m}. \] 

The system for first moments is relatively straightforward: first, the means $M_{m|m}(t)$ where $m = 1, \ldots, \mathcal{M}$ are simply solutions to pure death equations, so that 
\[M_{m|m}(t) = e^{\kappa_{mm} t} = e^{-\mu_m t}. \]
These solutions can now be substituted into simple first moment equations conditional on beginning with a marked progenitor: from \eqref{eq:genfirst}, these equations are given by
\[ \frac{\partial}{ \partial t} M_{ m|a}(t) = \kappa_{aa} M_{m|a}(t) + \indfun{a \rightarrow m} \kappa_{am} M_{m|m}(t), \]
and upon rearrangement are of the general form 
\begin{equation}\label{eq:intfactor} \frac{d}{dt} M_{m | a}(t) + P(t) M_{m| a}(t) =  Q(t) . \end{equation}
Such a differential equation can be solved using the integrating factor method, multiplying both sides $e^{\int P(t) dt} $ and rearranging for $M_{m | a}(t)$. Solving, we obtain
\[ M_{m|a}(t) = \indfun{a \rightarrow m} \frac{\kappa_{am}}{\kappa_{aa} - \kappa_{mm}} \left( e^{\kappa_{aa}t} - e^{\kappa_{mm}t} \right)  \quad = \quad
 \indfun{a \rightarrow m} \frac{\nu_m}{\mu_m - \mu_a} \left( e^{-\mu_a t} - e^{-\mu_m t}  \right) .\]
These expressions are intuitive: a higher rate of differentiation $\nu_m$ leads to an increase in the mean population of type $m$ cells, while a larger death rate $\mu_m$ relative to the death rate of progenitors $\mu_a$ producing the type $m$ cells decreases their mean population.
Next, \eqref{eq:genfirst} again gives us mean equations conditional on beginning with one marked HSC:
\[ \frac{\partial}{ \partial t} M_{ m|0}(t) = \kappa_{00} M_{m|0}(t) + \sum_{a \in \mathcal{A}} \indfun{a \rightarrow m} \kappa_{0a} M_{m|a}(t), \]
which clearly is also of the form \eqref{eq:intfactor}. Thus, we can plug in the solutions we've obtained for $M_{m|a}(t)$ and solve the system using the same technique, yielding
\begin{align*}
M_{m|0}(t) &= e^{\kappa_{00} t} \sum_{a \in \mathcal{A}} \indfun{a \rightarrow m} \frac{ \kappa_{0a} \kappa_{am} }{\kappa_{aa} - \kappa_{mm} } \left( \frac{ e^{(\kappa_{aa} - \kappa_{00})t} - 1 }{\kappa_{aa} - \kappa_{00}} - \frac{ e^{(\kappa_{mm} - \kappa_{00})t} - 1 }{\kappa_{mm} - \kappa_{00}} \right) \\
 &= e^{(\lambda - \sum_a \nu_a) t} \sum_{a \in \mathcal{A}} \indfun{a \rightarrow m} \frac{ \nu_a \nu_m }{\mu_m  -\mu_a } \left( \frac{ e^{(( \sum_a \nu_a) -\mu_a - \lambda )t} - 1 }{( \sum_a \nu_a) -\mu_a - \lambda} - \frac{ e^{(( \sum_a \nu_a) -\mu_m - \lambda)t} - 1 }{( \sum_a \nu_a) -\mu_m - \lambda } \right).
\end{align*}
These expressions characterize the mean behavior of the system, and furthermore may now be used toward solving for the second moments. We introduce for simplicity the additional notation $\kappa_{i,jk} := \frac{\partial^2 u_i}{\partial s_j \partial s_k} \bigm|_\mb{s=1}$; for instance, $\kappa_{0,00} = 2 \lambda$. Further, the equations $U_{mm|m}(t) = \kappa_{mm} U_{mm|m}(t)$, and together with the initial condition are only satisfied by the trivial solution $U_{mm|m}(t)=0$ for all final types $m$. Now, many terms in equation \eqref{eq:gensecond} have zero contribution, and the remaining equations in the system can be simplified to yield 
\begin{align*}
\frac{d}{dt} U_{mn| a}(t) &=  \indfun{a \rightarrow m} \indfun{a \rightarrow n} \left(\frac{ \partial u_a}{\partial s_a} \frac{\partial^2 \phi_a}{\partial s_m \partial s_n} + \frac{\partial^2 u_a}{\partial s_a \partial s_m} \frac{ \partial \phi_a}{\partial s_n}\frac{\partial \phi_m}{ \partial s_m} + \frac{\partial^2 u_a}{\partial s_a \partial s_n} \frac{ \partial \phi_a}{\partial s_m}\frac{\partial \phi_n}{ \partial s_n}  \right) \\
&=  \indfun{a \rightarrow m}\indfun{a \rightarrow n} \left( \kappa_{aa} U_{mn|a} + \kappa_{a,am} M_{n|a} M_{m|m} + \kappa_{a,an} M_{m|a} M_{n|n} \right) \quad \quad , \end{align*}

\begin{align*}
\frac{d}{dt} U_{mn| 0}(t) &= \left( \frac{ \partial u_0}{\partial s_0} \frac{ \partial^2 \phi_0 } {\partial s_m s_n} + 2 \frac{\partial^2 u_0}{\partial s_0^2} \frac{\partial \phi_0}{\partial s_m} \frac{\partial \phi_0}{\partial s_n} + \sum_{a \in \mathcal{A}} \indfun{ a \rightarrow m } \indfun{a \rightarrow n} \frac{\partial u_0}{\partial s_a} \frac{ \partial^2 \phi_a}{\partial s_m s_n} \right) \Bigm|_{\mb{s=1}} \\
&= \kappa_{00} U_{mn|0} + 2 \kappa_{0,00} M_{m|0} M_{n|0} + \sum_{a \in \mathcal{A}} \indfun{ a \rightarrow m } \indfun{a \rightarrow n}  \kappa_{0a} U_{mn|a} .
\end{align*}
Similarly,
\begin{align*}
\frac{d}{dt} U_{mm| a}(t) &=  \indfun{a \rightarrow m} \left(\frac{ \partial u_a}{\partial s_a} \frac{\partial^2 \phi_a}{\partial s_m ^2 } + 2 \frac{\partial^2 u_a}{\partial s_a \partial s_m} \frac{ \partial \phi_a}{\partial s_m}\frac{\partial \phi_m}{ \partial s_m} + 0 \right)  \\
&= \indfun{a \rightarrow m} \left( \kappa_{aa} U_{mm|a} + 2 \kappa_{a,am} M_{m|a} M_{m|m} \right), 
\end{align*}

\begin{align*}
\frac{d}{dt} U_{mm| 0}(t) &= \bigg[ \frac{ \partial u_0}{\partial s_0} \frac{ \partial^2 \phi_0 } {\partial s_m^2} + \frac{\partial^2 u_0}{\partial s_0^2} \left(\frac{\partial \phi_0}{\partial s_m}\right) ^2  + \sum_{a \in \mathcal{A}} \indfun{ a \rightarrow m } \frac{\partial u_0}{\partial s_a} \frac{ \partial^2 \phi_a}{\partial s_m^2 } \bigg] \Bigm|_{\mb{s=1}} \\
&= \kappa_{00} U_{mm|0} + \kappa_{0,00} M_{m|0}^2  + \sum_{a \in \mathcal{A}} \indfun{ a \rightarrow m } \kappa_{0a} U_{mm|a} .
\end{align*}
Since we already have expressions for the means $M_{\cdot | \cdot}$, these equations $U_{\cdot | a}(t)$ each become a first order linear ODE and can now each be solved individually. Indeed, they again take the form \eqref{eq:intfactor}, and we find
\begin{align*}
U_{mm|a}(t) &= \indfun{a \rightarrow m} e^{ \kappa_{aa} t} \int_{0}^t 2 \cdot e^{-\kappa_{aa} x} \kappa_{a,am} M_{m|a}(x) M_{m|m}(x) \, dx , \\
U_{mn|a}(t) &= \indfun{a \rightarrow m}\indfun{a \rightarrow n} e^{ \kappa_{aa} t} \int_{0}^t  e^{-\kappa_{aa} x} \left( \kappa_{a,am} M_{n|a}(x) M_{m|m}(x) + \kappa_{a,an} M_{m|a}(x) M_{n|n}(x) \right) \, dx .
\end{align*}
Replacing $\kappa_{\cdot}$ with explicit rates, we integrate and simplify these expressions to obtain
\begin{align*}
 U_{mm|a}(t) &= \indfun{a \rightarrow m}  \frac{2 \nu_m^2}{\mu_m - \mu_a} e^{-\mu_a t} \bigg[ \frac{\mu_a - \mu_m}{\mu_m( \mu_a - 2\mu_m)} - \frac{e^{-\mu_m t} }{\mu_m} - \frac{e^{(\mu_a - 2\mu_m)t}}{\mu_a - 2\mu_m} \bigg] \\
 U_{mn|a}(t) &= \indfun{a \rightarrow m}\indfun{a \rightarrow n} \bigg\{ \frac{\nu_m \nu_n}{\mu_n - \mu_a} e^{-\mu_a t} \bigg[ \frac{\mu_a - \mu_n}{\mu_m( \mu_a - \mu_m - \mu_n)} - \frac{e^{-\mu_m t} }{\mu_m} - \frac{e^{(\mu_a - \mu_m - \mu_n)t}}{\mu_a - \mu_m - \mu_n} \bigg]  \\
 & \quad + \frac{\nu_m \nu_n}{\mu_m - \mu_a} e^{-\mu_a t} \bigg[ \frac{\mu_a - \mu_m}{\mu_n( \mu_a - \mu_m - \mu_n)} - \frac{e^{-\mu_n t} }{\mu_n} - \frac{e^{(\mu_a - \mu_m - \mu_n)t}}{\mu_a - \mu_m - \mu_n} \bigg] \bigg\} .
\end{align*}
It is worth noting here that the product $\indfun{a \rightarrow m}\indfun{a \rightarrow n}$ is zero for any pair of types $m, n$ not descended from the same progenitor type, which may occur in models with specialized oligopotent progenitors. Recall that $\text{Cov}[X_m(t), X_n(t) | \mb{X}(0) = \mb{e}_a) ] = U_{mn|a}(t) - M_{m|a}(t) M_{n|a}(t)$. We see that in this case, $U_{mn}$ becomes zero, leading to lower values of the model-based covariance. In particular, $\text{Cov}[X_m(t), X_n(t) | \mb{X}(0) = \mb{e}_a) ]$ may be negative when $m, n$ do not share a progenitor type. This gives some intuition on the substantial effect of progenitor structure, which becomes apparent in the results presented in Section \ref{sec:results}.

Finally, we plug in these solutions into the differential equations beginning with an HSC governing $U_{\cdot | 0}(t)$, which now take on the same general form and again can be solved by the integrating factor method:
\begin{align*}
U_{mn|0}(t) &=  e^{ \kappa_{00} t} \int_{0}^t  e^{-\kappa_{00} x} \left( \kappa_{0,00} M_{n|0}(x) M_{m|0}(x) + \sum_{a \in \mathcal{A}} \indfun{a \rightarrow m}\indfun{a \rightarrow n} \kappa_{0a} U_{mn|a}(x) \right) \, dx  , \\
U_{mm|0}(t) &=  e^{ \kappa_{00} t} \int_{0}^t  e^{-\kappa_{00} x} \left( \kappa_{0,00} M_{m|0}^2(x) + \sum_{a \in \mathcal{A}} \indfun{a \rightarrow m} \kappa_{0a} U_{mm|a}(x) \right) \, dx . 
\end{align*}
These integrals have closed form solutions since their integrands only differ from the previous set of equations by additional exponentials contributed from the $U_{\cdot|a}(t)$ expressions. We omit the integrated forms in the general case for brevity, but remark that while they appear lengthy, they are comprised of simple terms and can be very efficiently computed within iterative algorithms. For completeness, we include the explicit solutions to the simplest model in the Appendix A.3. 

\subsection{Marginalized Moments}
With closed form moment expressions in hand, we can readily recover variance and covariance expressions for mature cells and thus calculate model-based correlations. For instance,
\[ \text{Cov}\left[ X_m(t), X_n(t)| \mb{X}(0) = \mb{e}_j \right] = U_{mn|j}(t) - M_{m|j}(t) M_{n|j}(t). \]
Because the initial state is uncertain, unconditional variances and covariances between mature types can be computed by marginalizing over the initial distribution vector $\boldsymbol\pi$. Derivations for all of the following expressions in this section are included in the Appendix A-3. We arrive at the marginal expressions by applying the law of total (co)variance:
\begin{align}
\V [ X_m(t) ] &= \sum_{k=1}^K \pi_k \E [ X_{m|k} ^2 ] - \sum_{k=1}^K \pi_k^2( \E[X_{m|k}])^2 ) - 2 \sum_{j > k} \pi_j \pi_k \E[ X_{m|j}] \E[ X_{m|k} ] \nonumber \\
&= \sum_{k=1}^K \pi_k [ U_{mm | k}(t) + M_{m | k}(t) ] - \pi_k^2 M_{m|k}(t)^2 - 2 \sum_{j > k} \pi_j \pi_k M_{m|k}(t) M_{m|j}(t). \label{eq:totalvar} \end{align}

\begin{align}
\cov{[ X_m(t), X_n(t)]} &= \sum_{k=1}^K \pi_k \E[ X_{m|k} X_{n|k}] - \sum_{k=1}^K \pi_k^2 \E[X_{m|k}] \E [X_{n|k} ] - \sum_{k\neq l} \pi_k \pi_l \E[X_{m|k}] \E[X_{n|l}] \nonumber \\ 
&= \sum_{k=1}^K \pi_k  U_{mn | k}(t)  -  \pi_k^2 M_{m|k}(t) M_{n|k}(t) -  \sum_{k \neq l} \pi_k \pi_l M_{m|k}(t) M_{n|l}(t) \label{eq:totalcov}. \end{align}
It remains to relate these expressions to the correlations between read counts $\psi_{mn}(\boldsymbol\theta; \mb{Y})$. 
Applying the law of total (co)variance again with respect to the multivariate hypergeometric sampling distribution, we obtain the following expressions:
\begin{align}
\cov{[Y_m(t), Y_n(t)]} &= \frac{b_m b_n}{B_m(t) B_n(t)} \cov{[X_m(t), X_n(t) ]} \label{eq:totvar}, \\ 
\text{Var}[Y_m(t)] &= \frac{b_m(B_m(t)-b_m)}{B_m(t)(B_m(t)-1)} \E[X_m(t)] - \frac{b_m(B_m(t)-b_m)}{B_m(t)^2(B_m(t)-1)}\E[ X_m ^2(t)] + \frac{b_m^2}{B_m(t)^2} \text{Var}[X_m(t)]. \nonumber
\end{align}
\vmadd{We note that the last set of variance and covariance expressions is an approximation, because we treated $B_m(t)$ as a constant for all $m$. In Section Appendix A.3 we provide a justification for this approximation. In our empirical evaluation of our moment-based estimators we did not observe any negative effects of this approximation.}

\subsection{Implementation}
We implemented these methods in the R package \texttt{branchCorr}, available at 
\url{https://github.com/jasonxu90/branchCorr}. Software includes algorithms to simulate and sample from this class of branching process models, to compute model-based moments given parameters, and to estimate parameters by optimizing the loss function objective. 
We provide a vignette that steps through smaller-scale reproductions of all simulations in this paper.

\section{Results}\label{sec:results}
\subsection{Simulation Study}\label{sec:sim}
We examine performance of the loss function estimator on simulated data, generated from several hematopoietic tree structures in our branching process framework. Specifically, we consider models with three or five mature types with varying progenitor structures displayed in Figure \ref{fig:allmodels}. Under each model, we simulate $400$ independent datasets, each consisting of $20,000$ realizations representing distinct barcode IDs, from the specified continuous-time branching process. Since we observe fairly constant \textit{in vivo} cell populations in the real data, true rates for simulating these processes were chosen such that summing over the $20,000$ barcodes, the total populations of each mature cell type are relatively constant after time $t=2$. Note that while total populations are stable, individual barcode trajectories display a range of heterogeneous behaviors, with many trajectories becoming extinct and others reaching very high counts. This reflects the barcode count behavior that we see  in the real data.

From each synthetic dataset, we then sample an \textit{observed dataset} according to the multivariate hypergeometric distribution, mimicking the noise from blood sampling. Observations are recorded at irregular times over a two year period similar to the span and frequency of the experimental sampling schedule. Parameter estimation is then performed on the simulated datasets.
\begin{figure}
\centering
\includegraphics[height=7.1cm]{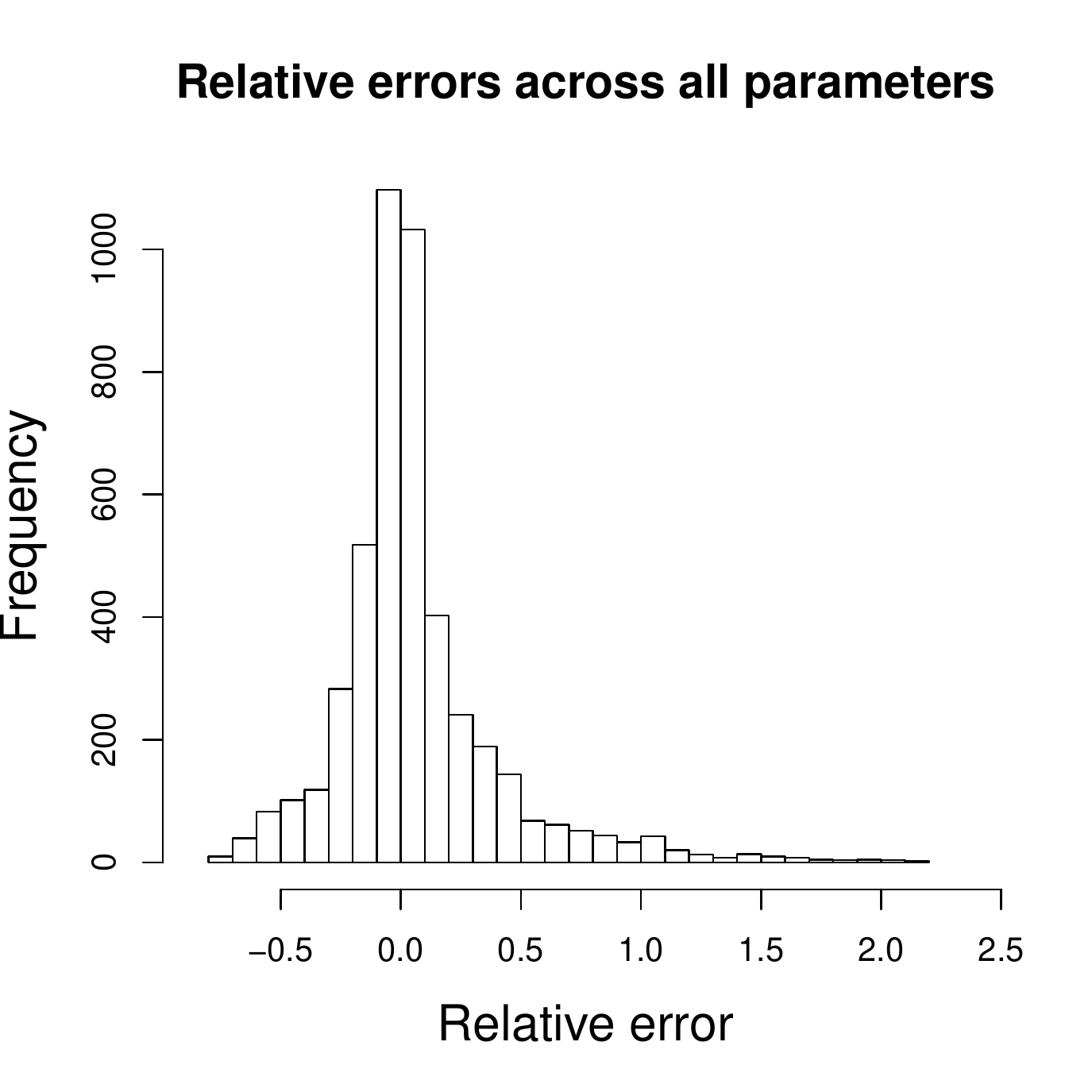} \includegraphics[height=7.1cm]{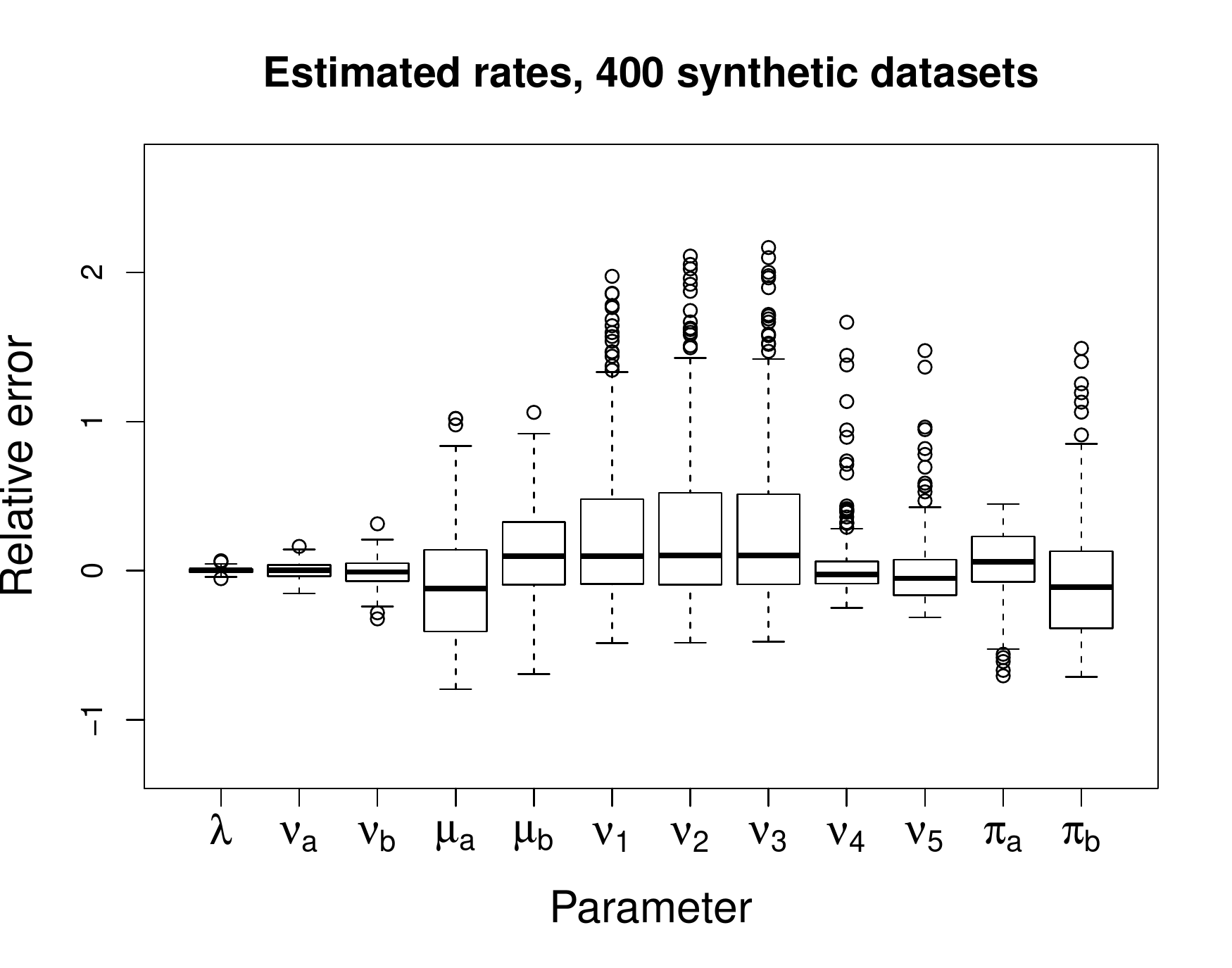} 

\caption[Performance of loss function estimator on simulated data]{ Performance of loss function estimator on synthetic data from model with five mature types and two progenitor types, i.e. model (c). While we see outlier influence, median estimates are accurate despite the parameter rich setting. Detailed medians, median absolute deviations, and standard errors corresponding to the plotted estimates are included in Appendix A-5. }
\label{fig:synthEst}
\end{figure}
To minimize the loss function, we use the general optimization implementation in R package \texttt{nlminb}. Optimization is performed over $250$ random restarts per dataset. We constrain rates to be non-negative, and include a log-barrier constraint to enforce that the overall growth of the HSC reserve is non-negative. The initial distribution vector is constrained to a probability simplex
via a multinomial logistic reparametrization;
see Appendix A-4 for details. Finally, we remark that optimization over all free parameters leads to mild identifiability problems---in particular, pairs of mature differentiation rates and death rates are often only identifiable up to a ratio. This is unsurprising: as correlations in the objective function are invariant to scale, we would expect some parameters to be distinguishable only up to a multiplicative constant. To remedy this, we choose to fix the death rates $\mu_i$ at their true value, supplying information that provides a sense of scale to infer all other parameters. Indeed, this is justifiable in practice, as mature cells are observable in the bloodstream, and information such as their average lifespan is available in the scientific literature.

\begin{figure}
\centering
\includegraphics[width = 0.496\textwidth]{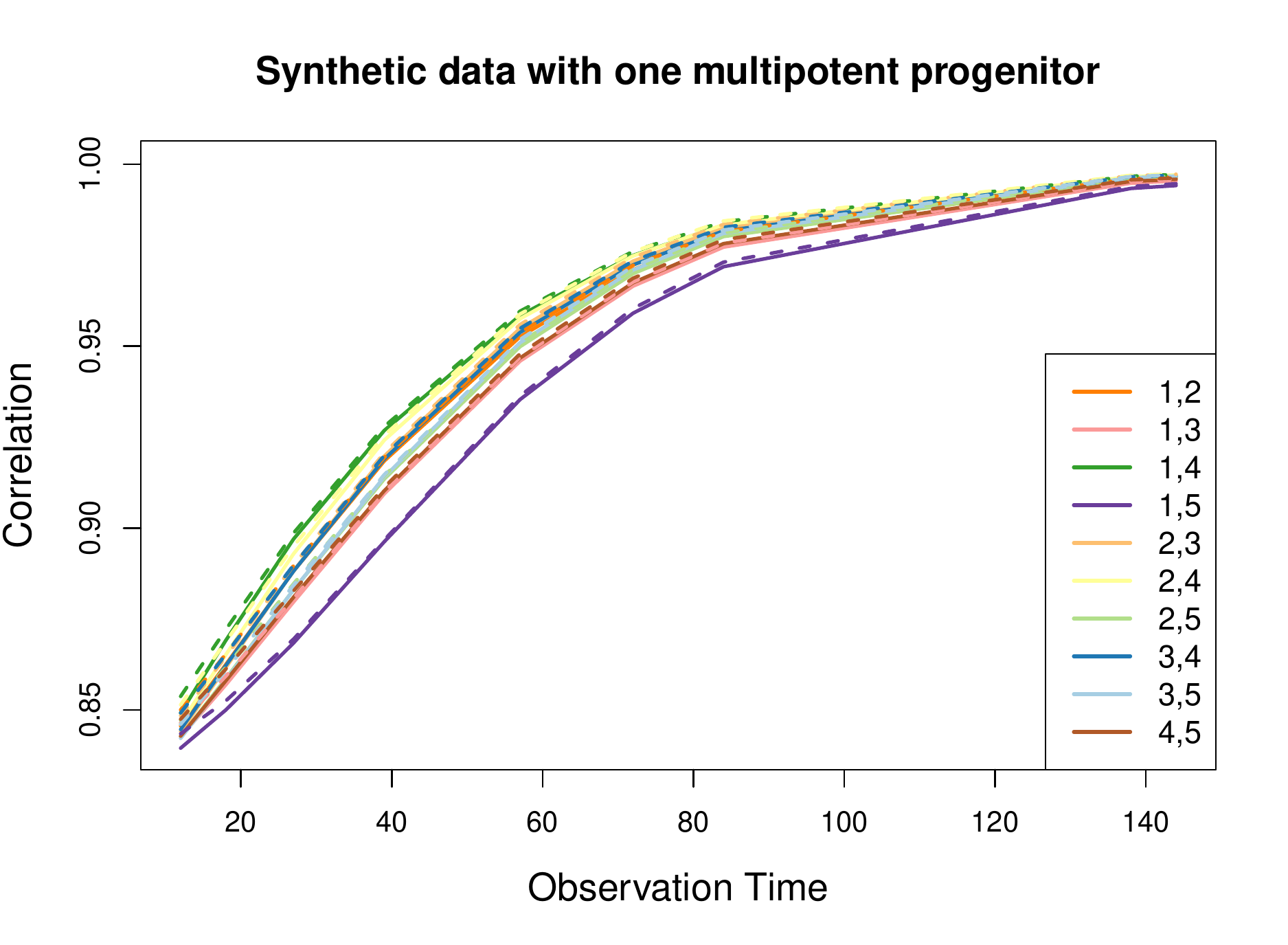} \includegraphics[width = 0.496\textwidth]{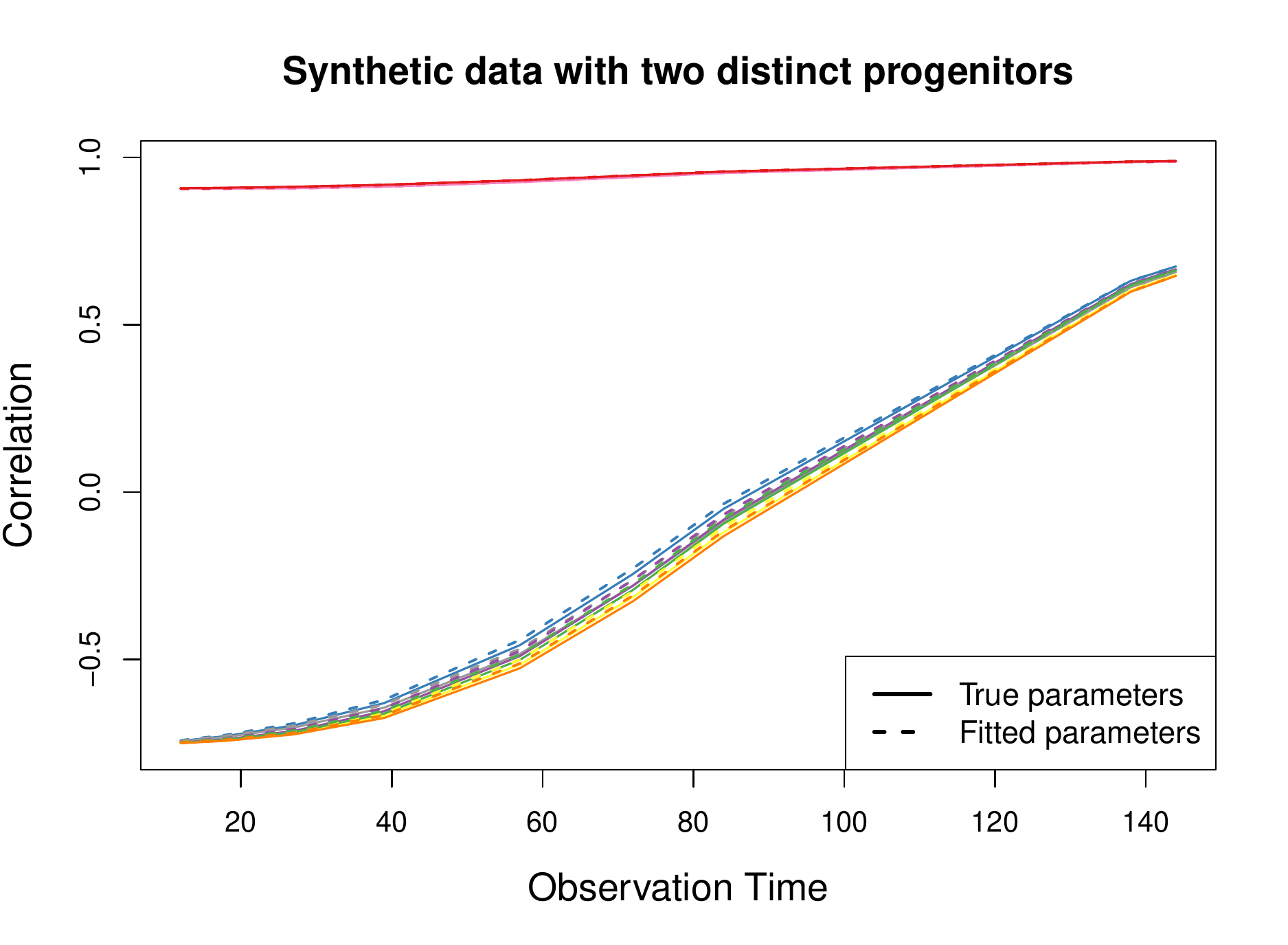}
\caption[Fitted correlation profiles on synthetic data]{ Pairwise correlation curves between five mature cell types descended from one common progenitor (left) or two distinct progenitors (right) calculated from our point estimate $\hat{\boldsymbol\theta}$. Solution curves from best fitting parameter estimates are almost indistinguishable from those corresponding to true parameters in both cases. Note that in the two-progenitor model, pairwise correlations among mature cell types display two clusters of behavior, and that negative correlations are possible. }
\label{fig:synthdata}
\end{figure}
 
Correlation profiles from estimated parameters corresponding to the results in the tables above are displayed in Figure \ref{fig:synthdata}. Visually, we see the fitted curves are very close to those corresponding to true parameters. We also note clear qualitative differences between models, with the two-progenitor model exhibiting two distinct groupings of correlation profiles, featuring low and negative correlations. 

\paragraph{Model misspecification}
In the following simulation experiments, we examine the performance of the estimator in under- and over-specified models. We do so by incorrectly assuming the data are generated from a model with one common progenitor or with three intermediate progenitors, and fitting these models to data simulated from the two-progenitor model (c) considered in the previous section. 
Figure \ref{fig:synthEst} shows that the median over relative errors $\frac{\hat\theta_i - \theta_i}{\theta_i}$ of each component in the estimated parameter vector $\hat{\boldsymbol\theta}$ is near zero, and we note the median value of the objective function \eqref{eq:momentsObjCorr} at convergence was $2.78 \times 10^{-4}$, with median absolute deviation $1.31 \times 10^{-4}$ and standard deviation $2.47 \times 10^{-4}$. 

The fitted correlation curves under misspecified progenitor structures are displayed in Figure \ref{fig:misPlots}, with detailed tables containing estimates again included in the Appendix A.5. We also examine the behavior when fitting a model with fewer types by ``lumping" similar mature cells together. To this end, we consider grouping mature types 2 and 3 together, and types 4 and 5 together, thus fitting a model with three total mature types, but with a progenitor structure consistent with the true model. Results in Figure \ref{fig:misPlots} suggest it is reasonable to group cells with shared lineages together, resulting in a much milder effect on model fit than progenitor structure misspecification. 
Such a grouping strategy can be important to avoiding overfitting a model to real data when some degree of model misspecification is inevitable, and is advantageous in settings where limited data suggest aggregation to reliably estimate fewer model parameters. 

\begin{figure}
\centering
\includegraphics[width = 1.01\textwidth]{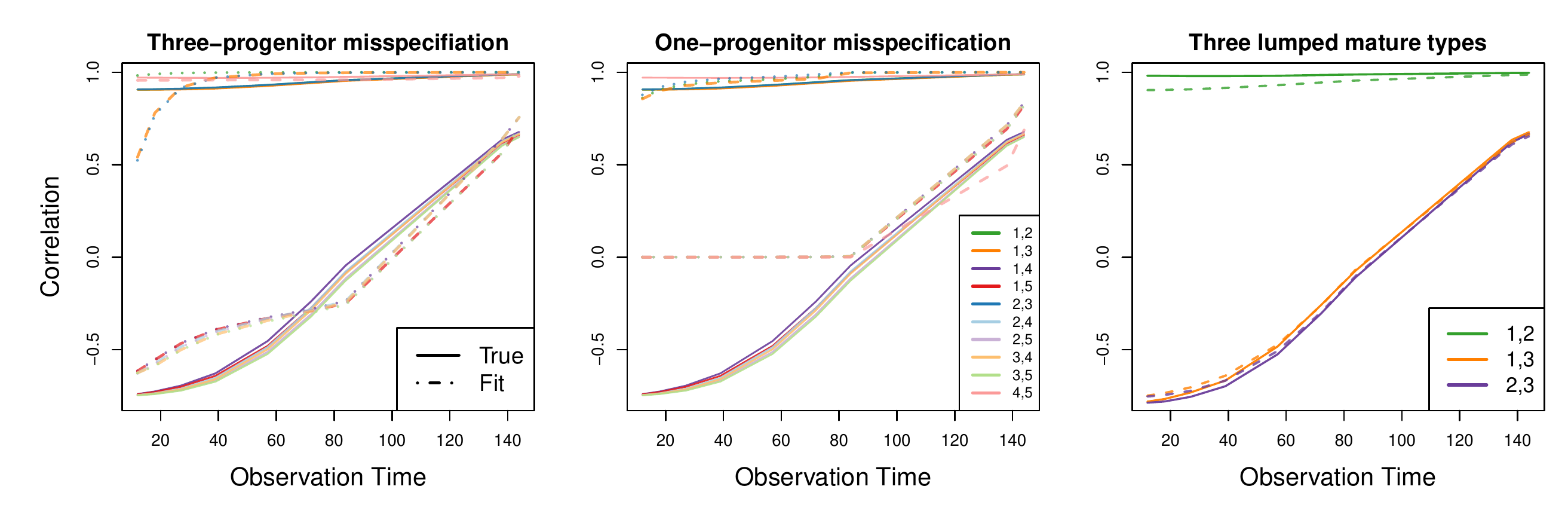}
\caption[Correlation curves under model misspecification]{Fitted correlation curves corresponding to misspecified model estimates. Data are generated from a true model with two distinct progenitors and the true correlation profiles are the same as those displayed in the right panel of Figure \ref{fig:synthdata}. While we see a generic lack of fit in the three-progenitor model, notice that specifying one common progenitor fails to exhibit negative correlations necessary to explain the data. On the other hand, ``lumping" mature cells but properly specifying progenitor structure results in reasonable performance, as evident in the rightmost panel.}
\label{fig:misPlots}
\end{figure}

\subsection{Cell Lineage Barcoding in Rhesus Macaques}\label{sec:realdata}
Having validated our method on data simulated from the model, we turn to analyze the lineage barcoding data from \cite{wu2014}. We consider barcoding data collected from a rhesus macaque over a $30$ month period following bone marrow transplantation. We consider only sampling times at which uncontaminated read data for each of the five cell types (granulocyte, monocyte, T, B, and Natural Killer) are available and, as in the original study, apply a filter so that we consider only lineages exceeding a threshold of at least $1000$ read counts at any time point. After these restrictions, our dataset consists of $9635$ unique barcode IDs, with read data available at eleven unevenly spaced sampling times.

As inputs to the loss function estimator, we fix death rates at biologically realistic parameters based on previous studies \citep{hellerstein1999, zhang2007, kaur2008}, reported below. Parameters of the multivariate hypergeometric sampling distribution are informed by circulating blood cell (CBC) data recorded at sampling times. These include $B_m(t)$, the total population of type $m$ cells in circulation at time $t$ across all barcodes, and $b_m$, the constant number of type $m$ cells in the sample at each observation time. 

\begin{figure}
\centering
\includegraphics[height=7.7cm]{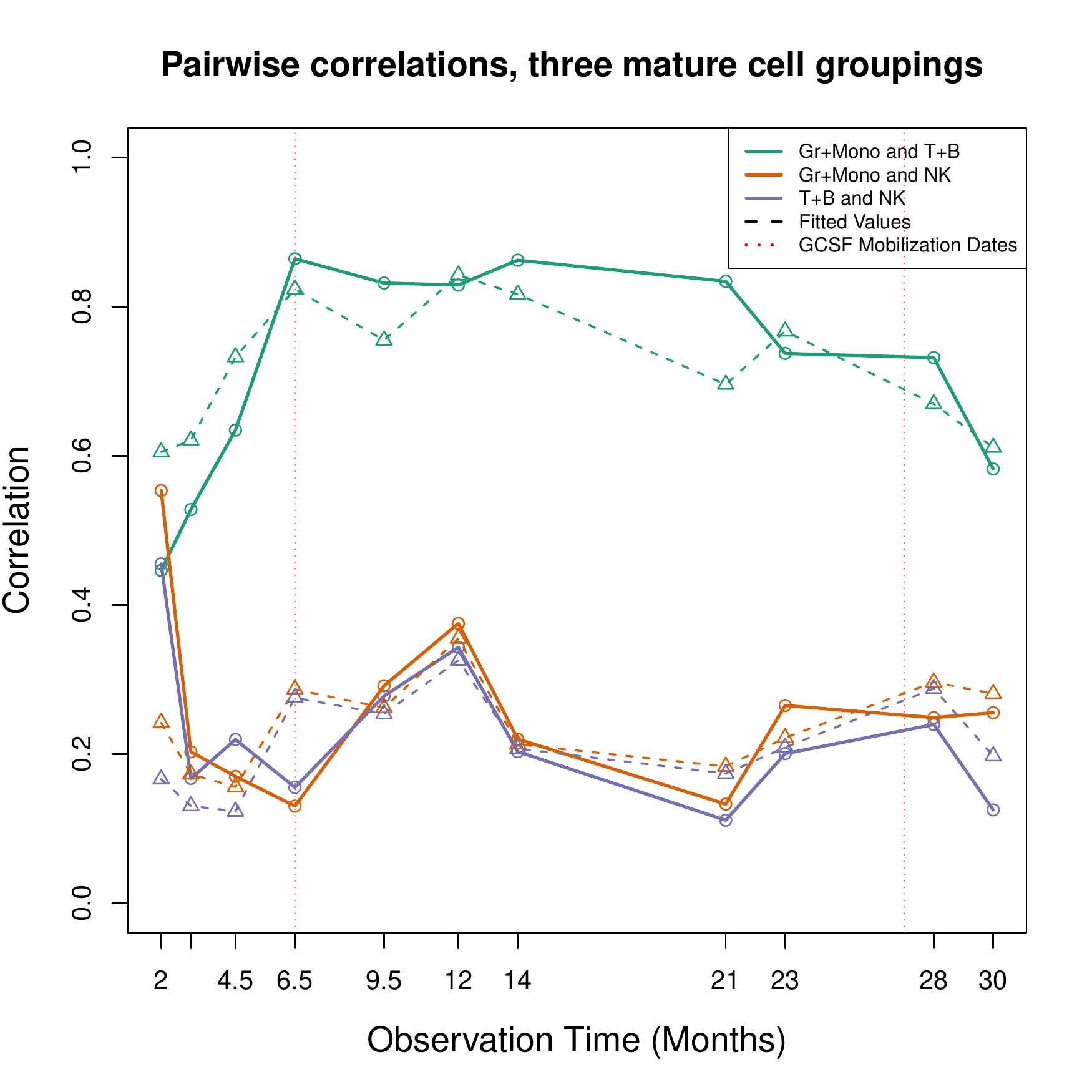} \hspace{-15pt} \includegraphics[height=7.7cm]{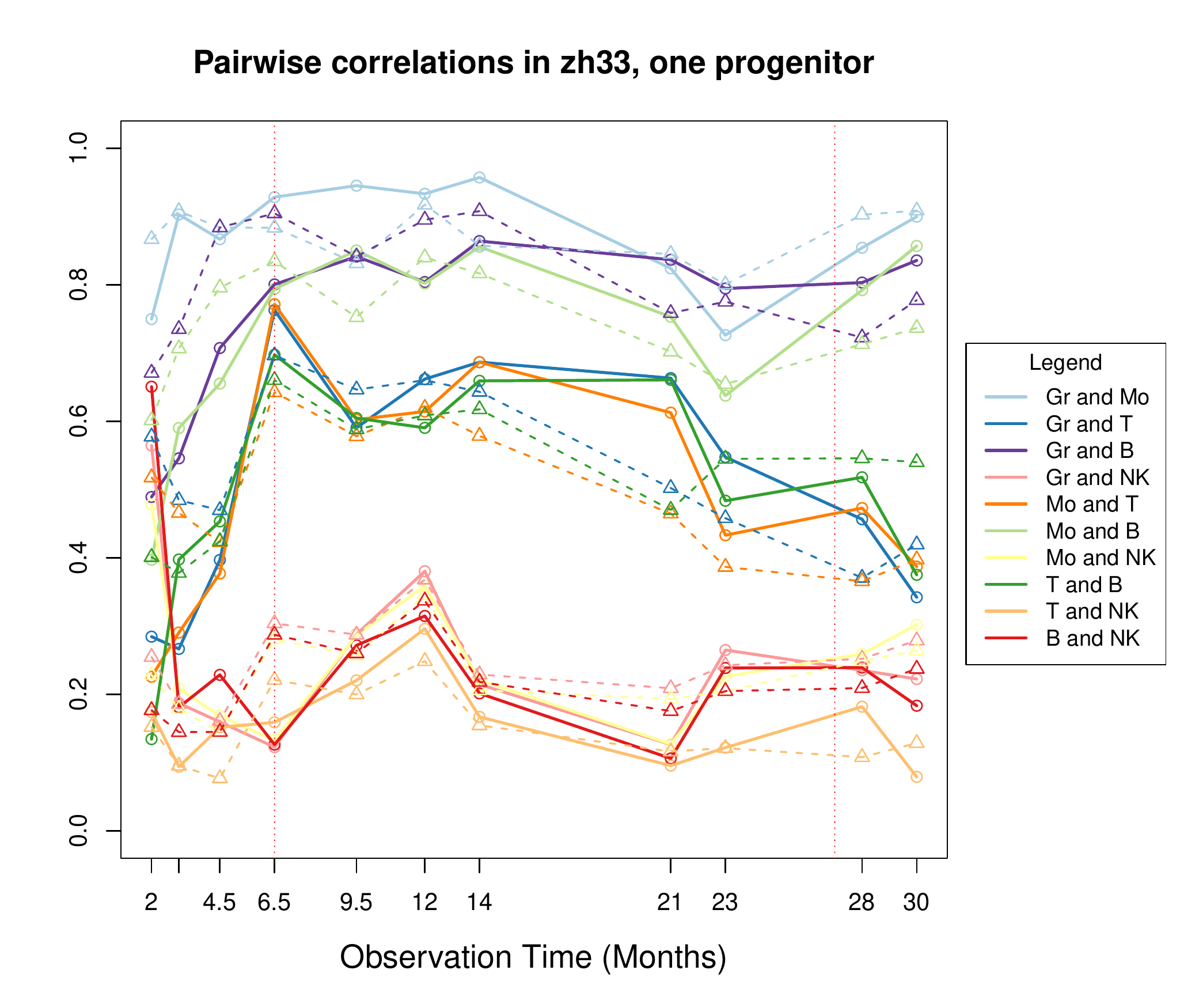}
\caption[Best-fit correlation curves in a one-progenitor model to rhesus macaque barcoding data]{Dashed lines depict fitted correlations to read data in models (a) and (b) assuming one early progenitor type. GCSF mobilization dates are marked by vertical red lines. Solid lines connect the empirical correlations. 
} \label{fig:realdata}
\end{figure}

We estimate the remaining rate parameters and initial barcoding distribution using the loss function estimator in all models displayed in Figure \ref{fig:allmodels}. Fitted pairwise correlation curves from  estimates obtained via loss function optimization with 2000 random restarts in models with one multipotent progenitor type are displayed in Figure \ref{fig:realdata}. There are three curves in model (a) with three mature types, and ten curves corresponding to possible pairs among the five mature types in model (b) plotted on the right. The empirical correlations from raw data are displayed as solid lines. On a qualitative level, there is visible separation into three clusters of correlation profiles among the five mature cell groups, consistent with the simpler lumped model (a). Notably, empirical correlations between NK cells and any other cell type are significantly lower than all other pairwise correlations. This supports the main result in the pilot clustering-based analysis in the original study \citep{wu2014}, reporting on distinctive NK lineage behavior, from a new perspective. In both plots, fitted curves  follow the shape of observed correlations over time, and we observe that the largest error occurs at the $6.5$ month sample, coinciding with the application of granulocyte-colony stimulating factor (GCSF), a technical intervention that perturbs normal hematopoiesis in the animal. 

\begin{figure}
\centering
\includegraphics[width=.90\textwidth]{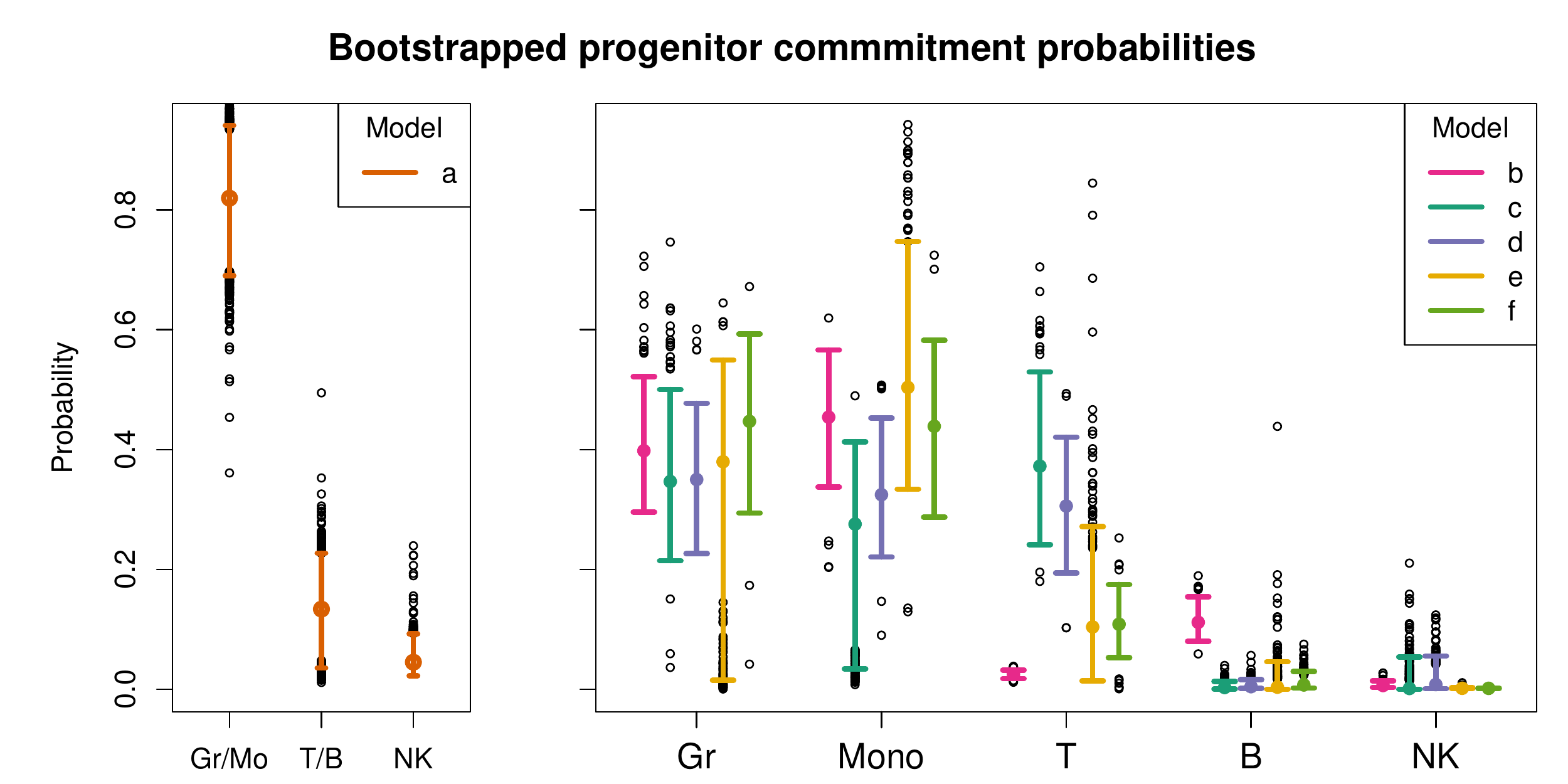} 
\caption[Comparison of fitted fate decision probabilities of progenitor commitment]{Comparison of fitted intermediate differentiation rates parametrized as fate decision probabilities. Displayed are the bootstrap estimates of normalized commitment rates to each mature type $i$, $\frac{\hat\nu_i}{\sum_j \hat\nu_j}$, in each model displayed in Figure \ref{fig:allmodels} (a)-(f) fitted to rhesus macaque data. } \label{fig:fatedecisions}
\end{figure}

\begin{figure}
\centering
\includegraphics[width=.6\textwidth]{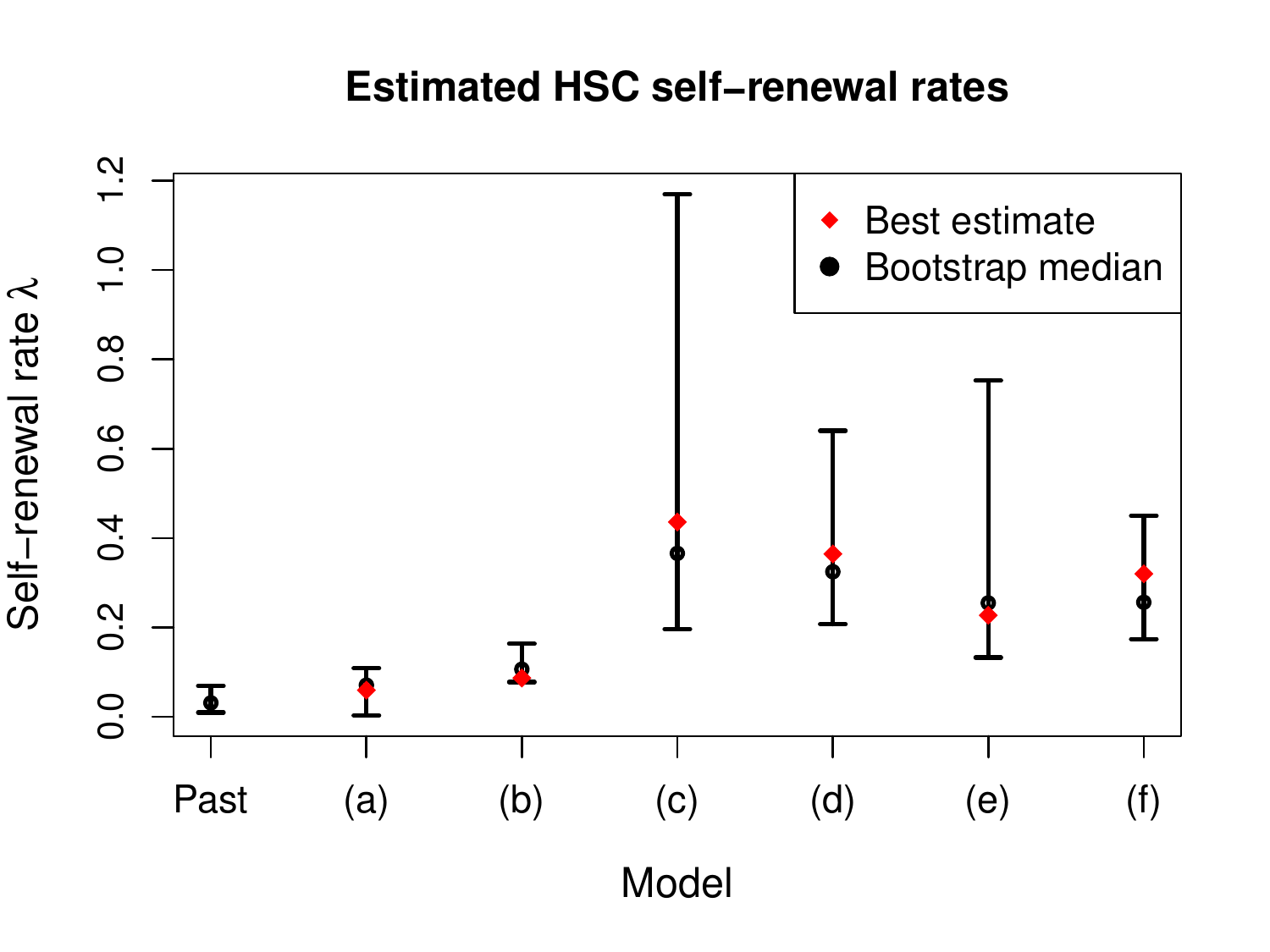} 
\caption[Comparison of fitted HSC self-renewal rates]{Comparison of fitted self-renewal rates $\hat\lambda$ and $95\%$ confidence intervals across all models displayed in Figure \ref{fig:allmodels} (a)-(f). Point estimates with lowest objective value (best estimates) are marked by red diamonds, while bootstrap confidence intervals and medians are plotted in black. The confidence interval around $\hat\lambda$ from model (a) overlaps with the interval obtained in previous telomere analyses focusing on HSC behavior in primates \citep{shepherd2007}, while the interval from model (b) is very close and in reasonable range. The other models, which do not feature a multipotent common progenitor, result in less biologically plausible estimates.} \label{fig:selfrenew}
\end{figure}

Next, we display a comparison of intermediate differentiation rates normalized as fate decision probabilities in Figure \ref{fig:fatedecisions} and fitted self-renewal rates in Figure \ref{fig:selfrenew} across models. The complete set of parameter estimates (used to generate fitted curves in Figure \ref{fig:realdata}) and their corresponding confidence intervals are reported in Appendix A-5. Confidence intervals are produced via $2500$ bootstrap replicate datasets. Nonparametric bootstrap resampling was performed over barcode IDs as well as over read count sampling, to account for both variation across stochastic realizations of barcode count time series and from sampling noise.  

Rate estimates are parametrized as number of events per five days: for instance, death rates $\boldsymbol\mu = (0.4, 0.04, 0.3)$ in the lumped model correspond to half-lives of about eight days among granulocytes and monocytes, three months for T and B cells, and two weeks in NK cells. In all models with five mature types, we fix death rates at $\boldsymbol\mu = (0.8, 0.3, 0.04, 0.08, 0.4)$. 

Previous studies of HSC dynamics in nonhuman primates based on telomere analysis \citep{shepherd2007} estimate the HSC self-renewal rate at once every 23 weeks, with 11-75 week range, corresponding to an estimate of $\widetilde{\lambda} = 0.0310$ with interval $(0.0095, 0.0649)$ when translated to our parametrization. As we see in Figure \ref{fig:selfrenew}, these findings coincide with our estimates and confidence intervals for $\hat\lambda$ in models with one multipotent progenitor. While other rates pertaining to intermediate cell stages and initial barcoding level are quantities that have not been previously estimated, our results suggest that granulocytes and monocytes are produced much more rapidly than T, B and NK cells and that individual progenitor cells are long-lived and can each produce thousands of these mature cells per day, both are biologically reasonable results. 
The initial barcoding percentage of HSCs is estimated at $13\%$ in model (a), depicted in Figure~\ref{fig:allmodels}. Since we experienced numerical instabilities while fitting models (b)--(f), we used model (a) estimate and fixed the total progenitor marking percentage at $87\%$ in these more complex models. However, estimates of $\hat{\boldsymbol{\pi}}$ in models with multiple progenitors lie on the boundary of the probability simplex, even when fixing $\hat\pi_0$ (see Supplementary Table \ref{tab:realdata}). Along with higher objective values and less biologically plausible parameters, these results suggest a poorer model fit, reminiscent of the behavior in the model misspecification experiments in Section \ref{sec:sim}.
We quantify this lack of fit by performing model selection via $5$-fold cross-validation (CV). 
We divide the dataset into five random subsets of equal size and fit each model to the training data consisting of the four of the subsets while holding one subset out as test data to assess predictive performance.
We then compute the objective function \eqref{eq:momentsObjCorr} using parameters obtained from the training data and empirical correlations computed using the test data.
These cross-validated objective function values for models (b)--(f), displayed in Table \ref{tab:cv}, are the average of the objectives evaluated across the five sets of training and test data. 
Model (a), not displayed in the table, achieves a CV objective value of $1.72$, but is not directly comparable as its loss function is comprised of fewer correlation terms since there are only three mature blood cell types. 
The CV objective value for the multipotent progenitor model (b) is noticeably lower, favoring this simple single progenitor model over more complex alternatives. 
In addition, models with oligopotent progenitors (c)--(f) visually fit the data worse than the multipotent progenitor model (b) when fitted and empirical correlations are plotted together (see Appendix A-5).


\begin{table}[htbp]
\centering
\begin{tabular}{l  c  c  c  c  c}
Model & (b) & (c) & (d) & (e) & (f) \\ 
\hline
Number of progenitors & 1 & 2 & 2 & 3 & 3  \\
CV Objective & 4.34 & 7.49 & 6.61 &  9.15 & 8.25   \\ 
\end{tabular}
\caption{Model selection via cross-validation with five folds, where the first row refers to specific models illustrated in Figure \ref{fig:allmodels}. The multipotent progenitor model (b)  results in a better objective value evaluated on held-out data than those with multiple oligopotent progenitors.  }
\label{tab:cv}
\end{table}


\section{Discussion}
We propose a novel modeling framework and parameter estimation procedure for analyzing hematopoietic lineage tracking experiments. To our knowledge, this is the first such method for fitting time series counts from cell lineage tracking data to continuous-time stochastic models featuring HSC, progenitor, and mature stages of cell development.
Detailed simulation studies show that the loss function estimator yields accurate inference when applied to data generated from this class of models. 
Our analysis of  \textit{in vivo} experimental data yields estimates of HSC self-renewal rates, intermediate cell differentiation rates, and progenitor death rates. 
We are the first to estimate most of these parameters in a large primate system.
Moreover, our methodology opens the door for statistically rigorous selection of models describing the hierarchy of hematopoietic cell specialization and differentiation.

\vmadd{
Our exploration of several models suggests that a model with one non-restricted multipotent progenitor provides a better fit to the data than models requiring an ordered hierarchical differentiation. 
This result may seem counterintuitive, but one needs to remember  that even though our models with multiple progenitors are more complex, they are also more restrictive in the sense that they include loss of lineage potential by limiting the types of mature cells that can descend from each distinct progenitor. 
If in reality progenitors never fully lose their potential to produce all mature cell types, this restriction leads to model misspecification. Indeed, recent studies dispute traditional assumptions about hematopoietic structures prescribing restricted differentiation pathways. For instance,  \citet{kawamoto2010} challenge the classical notion of a specialized myeloid progenitor, showing that lymphocyte progenitors (i.e. T, B, NK) can also give rise to myeloid cells (Gr and Mono). Recent \textit{in vitro} studies of human hematopoiesis suggest multipotence of early progenitors \citep{notta2015} may only occur in mature systems, and argue that oligopotent behavior is only observed in early stages of development. In light of this and other recent studies, our model selection results are supportive of emerging experimental data \citep{Velten2017}.}
\par
Several limitations remain when modeling hematopoiesis as a Markov branching process. The assumptions of linearity and rate homogeneity imply a possibility of unlimited growth, and extending analysis to allow for nonlinear regulatory behavior as the system grows near a carrying capacity is merited. Similarly, the Markov assumption may be relaxed to include arbitrary lifespan distributions---age-dependent processes are one example falling under this model relaxation, and have been applied to analyzing stress erythropoiesis in recent studies \citep{hyrien2015}. Further phenomena such as immigration or emigration in a random environment may be considered in future studies: for instance, it is known some cells in the peripheral bloodstream move in and out of tissue. While such extensions are mathematically challenging, they are straightforward to simulate, and various forward simulation approaches or approximate methods such as approximate Bayesian computation (ABC) \citep{marjoram2003,toni2009} may provide a viable alternative. Indeed, a Bayesian framework would further allow existing prior information available from previous studies about average lifespans of mature blood cells to be incorporated without fixing some of the model parameters.
\par
Our fully generative framework and accompanying estimator immediately enable simulation studies and sensitivity analyses, and can be adapted to developing model selection tools. The larger scientific problem of inferring the most likely lineage differentiation pathway structure directly translates to the statistical problem of model selection. Many model selection approaches essentially build on parameter estimation techniques, balancing model complexity and goodness of fit by penalizing the number of model parameters. 
While we perform model selection by loss function cross-validation, future work can investigate various penalization strategies applied to this class of models \citep{tibshirani1996, fan2001}, or with shrinkage priors in a Bayesian setting \citep{Park2008, Griffin2013}. Model selection using ABC, a well-studied and active area of research \citep{toni2009, liepe2014, Pudlo2016}, is also applicable to our modeling framework.
\par
Finally, the class of models we consider and derivations for their moment expressions are general in that an arbitrary number of intermediate progenitors and mature cell types can be specified. Nonetheless, these models have several limitations. 
First, we feature three stages of cell development in our model, and future work may extend this to include additional stages. 
Second, our assumptions only allow for each mature cell to be descended from one progenitor type, which limits the ability to investigate fully connected and nested models. Nonetheless, we have enabled parameter estimation in much more detailed models than previous statistical studies, while accounting for missing information and experimental noise. These models commonly arise in related fields such as chemical kinetics, oncology, population ecology, and epidemiology, and our methodology contributes broadly to the statistical toolbox for inference in partially observed stochastic processes, a rich area of research that still faces significant challenges. 

\section*{Acknowledgements}
VNM was supported by the NIH grants R01-AI107034 and U54-GM111274. JX was supported by the NSF MSPRF. CED, SK and CW were supported by the intramural program of the National Heart, Lung and Blood Institute. The authors thank Jon Wellner and Jon Wakefield for helpful discussions and guidance about GMM and $M$-estimation.

\bibliography{../../hema}

\newpage
\setcounter{table}{0}
\renewcommand{\thetable}{A-\arabic{table}}
\renewcommand{\thefigure}{A-\arabic{figure}}
\renewcommand{\thesection}{A-\arabic{section}}

\renewcommand{\theequation}{A-\arabic{equation}}
\setcounter{equation}{0}
\setcounter{section}{0}
\setcounter{figure}{0}

\begin{center}
\LARGE{Appendix} \end{center}

\section{Process specification using chemical kinetics notation}
In the chemical kinetics literature, the effect of events, or reactions, is often represented by the difference between system states before and after the occurrence of said reaction. In such systems, it is often the case that this difference is independent from the current system state. While this is not the case for branching processes, we describe how the probabilistic behavior of our models can be equivalently expressed in this representation, which may be more familiar to some audiences. 

Consider again the model in Figure 2(a) with three mature types. Let the current state of the system be denoted $\mb{x} = (n_0, n_a, n_1, n_2, n_3)$ as in the main text, and denote the components of another state $\mb{y} = (m_0, m_a, m_1, m_2, m_3)$. Then over an arbitrarily small interval of time $h$, the probability of moving from $\mb{x}$ to $\mb{y}$ is given by
\begin{equation*}
    \text{Pr}_{\mb{x}, \mb{y}}(h)=
    \begin{cases}
      n_0 \lambda h + o(h), & \quad  \mb{y} - \mb{x} = (1,0,0,0,0) \\
      n_0 \nu_a h + o(h), & \quad  \mb{y} - \mb{x} = (-1,1,0,0,0) \\
      n_a \nu_1 h + o(h), & \quad  \mb{y} - \mb{x} = (0,0,1,0,0) \\
      n_a \nu_2 h + o(h), & \quad  \mb{y} - \mb{x} = (0,0,0,1,0) \\
      n_a \nu_3 h + o(h), & \quad  \mb{y} - \mb{x} = (0,0,0,0,1) \\
      n_a \mu_a h + o(h), & \quad  \mb{y} - \mb{x} = (0,-1,0,0,0) \\
      n_1 \mu_1 h + o(h), & \quad  \mb{y} - \mb{x} = (0,0,-1,0,0) \\
      n_2 \mu_2 h + o(h), & \quad  \mb{y} - \mb{x} = (0,0,0,-1,0) \\
      n_3 \mu_3 h + o(h), & \quad  \mb{y} - \mb{x} = (0,0,0,0,-1) \\
      0, & \quad \text{otherwise}
    \end{cases}
 \end{equation*}
We see the similarity between this and the instantaneous rate vector notation for specifying a branching process.

\section{Consistency of loss function estimator}
There are several proofs of consistency and asymptotic normality for $M$-estimators or GMM estimators \citep{hansen1982, pakes1989, van2000} under various mild regularity assumptions. Consistency of our estimator follows from  Theorem 3.1 in \citep{pakes1989}. The result states that the vector $\hat{\boldsymbol\theta}_N$ which minimizes the norm $\lVert \mb{G}(\cdot) \rVert$ of a random, vector-valued function is consistent under the following conditions:
\begin{enumerate}[(i)]
\item[A1.]$ \displaystyle \lVert \mb{G}_N ( \hat{\boldsymbol\theta}_N) \rVert \leq o_p(1) + \inf_{\boldsymbol\theta \in \boldsymbol\Theta} \lVert \mb{G}_N ( \boldsymbol\theta) \rVert$,
\item[A2.] $\displaystyle \mb{G}_N (  \boldsymbol\theta_0)  = o_p(1)$, 
\item[A3.] $\displaystyle \sup_{ \lVert \boldsymbol\theta - \boldsymbol\theta_0 \rVert > \delta} \lVert \mb{G}_N ( \boldsymbol\theta) \rVert ^{-1} = O_p(1) \qquad$ for each $\delta>0$.
\end{enumerate}  
Heres $\boldsymbol\theta_0$ denotes the true data-generating parameters and is assumed to provide a global minimum to $\mb{G}$. A set of random variables $Z_n = o_p(1)$ if $Z_n$ converges to zero in probability, while $Z_n = O_p(1)$ if the set is stochastically bounded, i.e. for any $\varepsilon >0$, there exists finite $M$ such that $\text{Pr}( \lvert Z_n \rvert > M ) < \varepsilon$ for all $n$.

The first condition restricts us to estimators $\hat{\boldsymbol\theta}_N$ that nearly minimize $\lVert \mb{G}_N(\cdot) \rVert$. Condition A2 requires that under the true value, $\mb{G}_N (\boldsymbol\theta_0)$ converges to zero, which together with A1 implies that $\mb{G}_N (\hat{\boldsymbol\theta}_N)$ must also approach zero. Finally, condition A3 is an identifiability assumption, stating that small values of $\lVert \mb{G}_N (\boldsymbol\theta) \rVert$ can only occur near $\boldsymbol\theta_0$; this now forces $\hat{\boldsymbol\theta}_N$ to approach $\boldsymbol\theta_0$. 

\section{Derivation of second moments}
Here we explicitly derive the second moments of the simplest instance in our class of branching models of hematopoiesis. The derivation considers a four-type model with one progenitor and two mature types (Figure \ref{fig:allmodels} (a) ignoring the third mature compartment). We also derive the marginalized moment expressions after incorporating the sampling distribution.

From applying the process rates to the Kolmogorov backward equations, we can write \textit{pseudo-generating functions} defined 
\begin{equation}\label{eq:pseudogen} u_i(s_1,s_2,s_3,s_4) = \sum_j \sum_k \sum_l \sum_m  a_i(j,k,l,m)s_1^j s_2^k s_3^l s_4^m  . \end{equation}
For the model depicted in Figure \ref{fig:allmodels} (b), these are given by
\begin{align*}
u_1 (s_1, s_2) &= \lambda s_1^2 + \nu_0 s_2 - (\lambda + \nu_0) s_1, \\
u_2 (s_2,s_3,s_4) &= \nu_1 s_2 s_3 + \nu_2 s_2 s_4 + \mu_0 - ( \mu_0 + \nu_1 + \nu_2 ) s_2, \\
u_3 (s_3) &= \mu_1 - \mu_1 s_3; \quad \quad 
u_4( s_4) = \mu_2 - \mu_2 s_4.
\end{align*}
Next, we can write the probability generating function (PGF) of the process, beginning with one type $1$ particle, which is related to the pseudo-generating function $u_1$ as follows:
\begin{align}
\phi_1(t; s_1, s_2, s_3, s_4) &= \E\left[ \prod_{j=1}^4 s_j^{X_j(t)} | \mb{X}(0) = (1,0,0,0) \right] \nonumber \\
&= \sum_{k=0}^\infty \sum_{l=0}^\infty \sum_{m=0}^\infty \sum_{n=0}^\infty \text{Pr}_{(1,0,0,0),(k,l,m,n)} s_1^k s_2^l s_3^m s_4^n \nonumber \\
&= \sum_{k=0}^\infty \sum_{l=0}^\infty \sum_{m=0}^\infty \sum_{n=0}^\infty \left[ \mathbf{1}_{ \{ k=1, l = m =n =0 \} } + a_1(k,l,m,n) t + o(t) \right] s_1^k s_2^l s_3^m s_4^n \nonumber \\
&= s_1 + u_1(s_1, s_2, s_3, s_4) t + o(t). \label{eq:pgfAppendix}
\end{align}
We may analogously define  $\phi_i$ for processes beginning with one type $i$ particle, for each $i=1, \ldots, 4$. We have from Equation \eqref{eq:pgfAppendix} the relation
\[ \frac{\partial}{\partial t} \phi_i(t, s_1, \ldots, s_4) = u_i (\phi_1(t, s_1, \ldots, s_4), \ldots, \phi_4(t, s_1, \ldots, s_4) ). \]

Now, let $M_{l|k}(t)$ denote the expected number of type $l$ cells at time $t$, given one initial type $k$ cell. From definition of $\phi_i$, we see that we can relate the probability generating functions to these first moments via partial differentiation:
\[ M_{l | k}(t) = \frac{ \partial}{\partial s_l} \phi_k(t, s_1, \ldots, s_4) |_{s_1 = s_2 = s_3 = s_4 = 1}. \]
Similarly, we may further differentiate the PGF to derive second moments used toward variance and covariance calculations. 
Define
\[ U_{kl|1}(t) = \E \left[ X_k (X_l - \indfun{k=l} ) | \mb{X}(0) = (1,0,0,0) \right], \] 
with $U_{kl|i}(t)$ defined analogously beginning with one type $i$ particle.
Then $U_{kl|j}(t) = \frac{\partial^2 \phi_j}{\partial s_k \partial s_l} \bigg|_{\mb{s} = 1}$, and
by the Fa{\`a} di Bruno formula,
\[ \frac{ \partial^3 \phi_i}{\partial t \partial s_j \partial s_k} = \sum_{m= 1}^4 \left(\frac{\partial u_i}{\partial \phi_m} \frac{ \partial^2 \phi_m}{\partial s_j \partial s_k}\right) + \sum_{m,n=1}^4 \left( \frac{\partial^2 u_i}{\partial \phi_m \partial \phi_n} \frac{\partial \phi_m}{\partial s_j} \frac{\partial \phi_k}{\partial s_k} \right). \]
This relation allows us to write a system of non-homogeneous, linear ordinary differential equations (ODEs) governing second order moments:
\begin{align*}
\frac{\partial}{\partial t} U_{33|1}(t) &= (\lambda - \nu_0) U_{33|1}(t) + \nu_0 U_{33|2}(t) + (2 \lambda) M_{3|1}^2(t), \\ 
\frac{\partial}{ \partial t} U_{44|1}(t) &= (\lambda - \nu_0)U_{44|1}(t) + \nu_0 U_{44|2}(t) + (2 \lambda) M_{4 | 1}^2(t),  \\
\frac{\partial}{\partial t} U_{34|1}(t) &= (\lambda - \nu_0)U_{34|1}(t) + \nu_0 U_{34|2}(t) + (2 \lambda) M_{3|1}(t) M_{4 | 1}(t), \\
\frac{\partial}{\partial t} U_{34|2}(t) &= -\mu_0 U_{34|2}(t) + \nu_1 M_{4|2}(t) M_{3|3}(t) + \nu_2 M_{3|2}(t) M_{4|4}(t), \\
\frac{\partial}{\partial t} U_{33|2}(t) &=  -\mu_0 U_{33|2}(t) + \nu_1 U_{33|3}(t) + 2 \nu_1 M_{3|2}(t) M_{3|3}(t), \\
\frac{\partial}{\partial t} U_{44|2}(t) &= -\mu_0 U_{44|2}(t) + \nu_2 U_{44|4}(t) + 2 \nu_2 M_{4|2}(t) M_{4|4}(t), \\
\frac{\partial}{\partial t} U_{33|3}(t) &= -\mu_1 U_{33|3}(t),\\
\frac{\partial}{\partial t} U_{44|4}(t) &= -\mu_2 U_{44|4}(t), 
\end{align*}
all with initial conditions $(\cdot)_{k,l}(0) = 0$. We immediately see that $U_{33|3}(t) = U_{44|4}(t) = 0$, and upon a series of solutions and substitutions, we successively solve the system of ODEs, yielding the following explicit solutions:
\footnotesize
\begin{align*}
U_{33|2}(t) &= 2 \frac{\nu_1^2}{(\mu_2 - \mu_0)} \bigg[ \frac{ e^{-(\mu_0 + \mu_2)t}}{\mu_2} - \frac{e^{-2 \mu_2 t}}{\mu_0 - 2 \mu_2} + \frac{(\mu_0 - \mu_2) e^{-\mu_0 t }}{\mu_2(\mu_0 - 2\mu_2)} \bigg], \\
U_{44|2}(t) &= 2 \frac{\nu_2^2}{(\mu_2 - \mu_0)} \bigg[ \frac{ e^{-(\mu_0 + \mu_2)t}}{\mu_2} - \frac{e^{-2 \mu_2 t}}{\mu_0 - 2 \mu_2} + \frac{(\mu_0 - \mu_2) e^{-\mu_0 t }}{\mu_2(\mu_0 - 2\mu_2)} \bigg], \\
U_{34|2}(t) &= \frac{\nu_1 \nu_2}{(\mu_2 - \mu_0)} \bigg[ \frac{ e^{-(\mu_0 + \mu_1)t}}{\mu_1} - \frac{e^{-(\mu_1 + \mu_2) t}}{\mu_0 - \mu_1 - \mu_2} + \frac{(\mu_0 - \mu_2) e^{-\mu_0 t }}{\mu_1(\mu_0 - \mu_1 - \mu_2)} \bigg]  \\
& \quad \quad + \frac{\nu_1 \nu_2}{(\mu_1 - \mu_0)} \bigg[ \frac{ e^{-(\mu_0 + \mu_2)t}}{\mu_2} - \frac{e^{-(\mu_1 + \mu_2) t}}{\mu_0 - \mu_1 - \mu_2} + \frac{(\mu_0 - \mu_2) e^{-\mu_0 t }}{\mu_2(\mu_0 - \mu_1 - \mu_2)} \bigg], \\
U_{33|1}(t) &= e^{(\lambda - \nu_0)t} \bigg\{ 2 \frac{\nu_0 \nu_1 ^2}{ \mu_1-\mu_0 } \bigg[ \frac{ (\mu_0-\mu_1) e^{(\nu_0-\lambda-\mu_0)t}}{\mu_1(\mu_0 - 2\mu_1)(\nu_0-\lambda-\mu_0)}  - \frac{e^{(\nu_0-\lambda-\mu_0-\mu_1)t}}{\mu_1(\nu_0-\lambda-\mu_0-\mu_1)} - \frac{e^{(\nu_0-\lambda -2\mu_1)t}}{(\mu_0 - 2\mu_1)(\nu_0-\lambda-2\mu_1)} \\
& \quad \quad \quad \quad \quad  + \frac{\mu_1-\mu_0}{\mu_1 (\mu_0-2\mu_1)(\nu_0-\lambda-\mu_0)}  + \frac{1}{\mu_1(\nu_0-\lambda-\mu_0-\mu_1)}+ \frac{1}{(\mu_0-2\mu_1)(\nu_0-\lambda-2\mu_1)} \bigg] \\
& \quad \quad + \frac{2 \lambda\nu_0^2 \nu_1^2}{ (\mu_1-\mu_0)^2} \bigg[ \frac{e^{(\nu_0-\lambda-2\mu_0)t)}}{(\nu_0-\lambda-\mu_0)^2(\nu_0-\lambda-2\mu_0)} - \frac{2 e^{(\nu_0-\lambda-\mu_0-\mu_1)t}}{(\nu_0-\lambda-\mu_0)(\nu_0-\lambda-\mu_1)(\nu_0-\lambda-\mu_0-\mu_1)}\\
& \quad \quad \quad + \frac{2 (\mu_0-\mu_1)e^{-\mu_0 t}}{\mu_0 (\nu_0-\lambda-\mu_1)(\nu_0-\lambda-\mu_0)^2} + \frac{e^{(\nu_0-\lambda-2\mu_1)t}}{(\nu_0-\lambda-\mu_1)^2 (\nu_0-\lambda-2\mu_1)} + \frac{2(\mu_1-\mu_0) e^{-\mu_1 t}}{\mu_1 (\nu_0-\lambda-\mu_1)^2 (\nu_0-\lambda-\mu_0)} \\
& \quad \quad \quad + \frac{(\mu_1-\mu_0)^2 e^{(\lambda-\nu_0)t}}{(\lambda-\nu_0)(\nu_0-\lambda-\mu_1)^2 (\nu_0-\lambda-\mu_0)^2} - \frac{1}{(\nu_0-\lambda-\mu_0)^2 (\nu_0-\lambda-2\mu_0)} \\
& \quad \quad \quad + \frac{2}{(\nu_0-\lambda-\mu_0)(\nu_0-\lambda-\mu_1)(\nu_0-\lambda-\mu_0-\mu_1)} - \frac{2(\mu_0-\mu_1)}{ \mu_0 (\nu_0-\lambda-\mu_1)(\nu_0-\lambda-\mu_0)^2} \\
&\quad \quad \quad - \frac{1}{(\nu_0-\lambda-\mu_1)^2 (\nu_0-\lambda-2\mu_1)} - \frac{ 2(\mu_1-\mu_0)}{\mu_1 (\nu_0-\lambda-\mu_1)^2 (\nu_0-\lambda-\mu_0)} \\
& \quad \quad \quad - \frac{(\mu_1-\mu_0)^2}{(\lambda-\nu_0)(\nu_0-\lambda-\mu_1)^2(\nu_0-\lambda-\mu_0)^2) } \bigg] \bigg\}, \end{align*}
\footnotesize
\begin{align*}
U_{44|1}(t) &= e^{(\lambda - \nu_0)t} \bigg\{ 2 \frac{\nu_0 \nu_2 ^2}{ \mu_2-\mu_0 } \bigg[ \frac{ (\mu_0-\mu_2) e^{(\nu_0-\lambda-\mu_0)t}}{\mu_2(\mu_0 - 2\mu_2)(\nu_0-\lambda-\mu_0)}  - \frac{e^{(\nu_0-\lambda-\mu_0-\mu_2)t}}{\mu_2(\nu_0-\lambda-\mu_0-\mu_2)} - \frac{e^{(\nu_0-\lambda -2\mu_2)t}}{(\mu_0 - 2\mu_2)(\nu_0-\lambda-2\mu_2)} \\
& \quad \quad \quad \quad \quad  + \frac{\mu_2-\mu_0}{\mu_2 (\mu_0-2\mu_2)(\nu_0-\lambda-\mu_0)}  + \frac{1}{\mu_2(\nu_0-\lambda-\mu_0-\mu_2)}+ \frac{1}{(\mu_0-2\mu_2)(\nu_0-\lambda-2\mu_2)} \bigg] \\
& \quad \quad + \frac{2 \lambda\nu_0^2 \nu_2^2}{ (\mu_2-\mu_0)^2} \bigg[ \frac{e^{(\nu_0-\lambda-2\mu_0)t)}}{(\nu_0-\lambda-\mu_0)^2(\nu_0-\lambda-2\mu_0)} - \frac{2 e^{(\nu_0-\lambda-\mu_0-\mu_2)t}}{(\nu_0-\lambda-\mu_0)(\nu_0-\lambda-\mu_2)(\nu_0-\lambda-\mu_0-\mu_2)} \\
& \quad \quad \quad + \frac{2 (\mu_0-\mu_2)e^{-\mu_0 t}}{\mu_0 (\nu_0-\lambda-\mu_2)(\nu_0-\lambda-\mu_0)^2} + \frac{e^{(\nu_0-\lambda-2\mu_2)t}}{(\nu_0-\lambda-\mu_2)^2 (\nu_0-\lambda-2\mu_2)} + \frac{2(\mu_2-\mu_0) e^{-\mu_2 t}}{\mu_2 (\nu_0-\lambda-\mu_2)^2 (\nu_0-\lambda-\mu_0)} \\
& \quad \quad \quad + \frac{(\mu_2-\mu_0)^2 e^{(\lambda-\nu_0)t}}{(\lambda-\nu_0)(\nu_0-\lambda-\mu_2)^2 (\nu_0-\lambda-\mu_0)^2} - \frac{1}{(\nu_0-\lambda-\mu_0)^2 (\nu_0-\lambda-2\mu_0)} \\
& \quad \quad \quad + \frac{2}{(\nu_0-\lambda-\mu_0)(\nu_0-\lambda-\mu_2)(\nu_0-\lambda-\mu_0-\mu_2)} - \frac{2(\mu_0-\mu_2)}{ \mu_0 (\nu_0-\lambda-\mu_2)(\nu_0-\lambda-\mu_0)^2} \\
&\quad \quad \quad - \frac{1}{(\nu_0-\lambda-\mu_2)^2 (\nu_0-\lambda-2\mu_2)} - \frac{ 2(\mu_2-\mu_0)}{\mu_2 (\nu_0-\lambda-\mu_2)^2 (\nu_0-\lambda-\mu_0)}  \\
& \quad \quad \quad - \frac{(\mu_2-\mu_0)^2}{(\lambda-\nu_0)(\nu_0-\lambda-\mu_2)^2(\nu_0-\lambda-\mu_0)^2) } \bigg] \bigg\}, \\
\end{align*}
\footnotesize
\begin{align*}
U_{34|1}(t) &= e^{(\lambda - \nu_0)t} \bigg\{ \frac{\nu_0 \nu_1 \nu_2}{\mu_2-\mu_0} \cdot \bigg[  \frac{(\mu_0-\mu_2) e^{ (\nu_0-\lambda-\mu_0) t }}{\mu_1 (\mu_0-\mu_1-\mu_2) (\nu_0-\lambda-\mu_0) } - \frac{e^{(\nu_0-\lambda-\mu_1-\mu_0)t}}{\mu_1(\nu_0-\lambda-\mu_1-\mu_0) } \\
& \quad \quad - \frac{ e^{(\nu_0-\lambda-\mu_1-\mu_2) t}}{ (\mu_0-\mu_1-\mu_2) (\nu_0-\lambda-\mu_1-\mu_2) } + \frac{\mu_2-\mu_0}{\mu_1(\mu_0-\mu_1-\mu_2) (\nu_0-\lambda-\mu_0)} \\
& \quad \quad + \frac{1}{\mu_1 (\nu_0-\lambda-\mu_1-\mu_0)} + \frac{1}{(\mu_0-\mu_1-\mu_2)(\nu_0-\lambda-\mu_1-\mu_2)} \bigg] \\
& \quad \quad + \frac{ \nu_0 \nu_1 \nu_2}{ \mu_1-\mu_0} \bigg[  \frac{(\mu_0-\mu_1) e^{ (\nu_0-\lambda-\mu_0)t}}{ \mu_2 (\mu_0-\mu_1-\mu_2) (\nu_0-\lambda-\mu_0)} - \frac{ e^{ (\nu_0 - \lambda - \mu_2 - \mu_0)t}}{ \mu_2 (\nu_0 - \lambda - \mu_2 - \mu_0) } \\
& \quad \quad -\frac{ e^{(\nu_0 - \lambda - \mu_1 - \mu_2)t}}{ (\mu_0-\mu_1-\mu_2) (\nu_0 - \lambda - \mu_1 - \mu_2) }  + \frac{ \mu_1-\mu_0}{ \mu_2 (\mu_0-\mu_1-\mu_2)(\nu_0-\lambda-\mu_0)} \\
& \quad \quad + \frac{1}{ \mu_2 (\nu_0-\lambda-\mu_2-\mu_0)}  + \frac{1}{ (\mu_0-\mu_1-\mu_2) (\nu_0-\lambda-\mu_1-\mu_2) } \bigg]  \\
& + \frac{ 2 \lambda \nu_0^2 \nu_1 \nu_2}{ (\mu_1-\mu_0) (\mu_2-\mu_0)} \cdot \bigg[ \frac{ e^{ (\nu_0-\lambda-2 \mu_0)t}}{  (\nu_0-\lambda-2\mu_0) (\nu_0-\lambda-\mu_0)^2 }  - \frac{e^{(\nu_0-\lambda-\mu_0-\mu_2)t}}{ (\nu_0-\lambda-\mu_0) (\nu_0-\lambda-\mu_2) (\nu_0-\lambda-\mu_0-\mu_2) } \\
 & \quad \quad  + \frac{ (\mu_0-\mu_2) e^{-\mu_0 t} }{ \mu_0 (\nu_0-\lambda-\mu_0)^2 (\nu_0-\lambda-\mu_2)} - \frac{ e^{ (\nu_0-\lambda-\mu_0-\mu_1) t}}{(\nu_0-\lambda-\mu_0) (\nu_0-\lambda-\mu_1) (\nu_0-\lambda-\mu_0-\mu_1)} \\
 & \quad \quad + \frac{  e^{ (\nu_0-\lambda-\mu_1-\mu_2) t} }{ (\nu_0-\lambda-\mu_1) (\nu_0-\lambda-\mu_2) (\nu_0-\lambda-\mu_1-\mu_2)}  + \frac{ (\mu_2-\mu_0) e^{ -\mu_1 t}}{ \mu_1(\nu_0-\lambda-\mu_1) (\nu_0-\lambda-\mu_2) (\nu_0-\lambda-\mu_0)} \\
 & + \frac{ (\mu_0-\mu_1) e^{ -\mu_0 t}}{ \mu_0 (\nu_0-\lambda-\mu_0)^2 (\nu_0-\lambda-\mu_1)} + \frac{ (\mu_1-\mu_0) e^{ -\mu_2 t} }{ \mu_2(\nu_0-\lambda-\mu_1) (\nu_0-\lambda-\mu_2) (\nu_0-\lambda-\mu_0)} \\
 &  \quad \quad + \frac{ (\mu_1-\mu_0) (\mu_2-\mu_0) e^{ (\lambda-\nu_0) t)}} { (\lambda-\nu_0) (\nu_0-\lambda-\mu_0)^2 (\nu_0-\lambda-\mu_1) (\nu_0-\lambda-\mu_2) } - \frac{1}{ (\nu_0-\lambda-\mu_0)^2 (\nu_0-\lambda-2 \mu_0)}  \\
 & + \frac{1}{ (\nu_0-\lambda-\mu_0)(\nu_0-\lambda-\mu_2)(\nu_0-\lambda-\mu_0-\mu_2)}  - \frac{\mu_0-\mu_2}{ \mu_0 (\nu_0-\lambda-\mu_0)^2 (\nu_0-\lambda-\mu_2)} \\
 & + \frac{1}{ (\nu_0-\lambda-\mu_0)(\nu_0-\lambda-\mu_1)(\nu_0-\lambda-\mu_0-\mu_1) } - \frac{1}{ (\nu_0-\lambda-\mu_1)(\nu_0-\lambda-\mu_2)(\nu_0-\lambda-\mu_1-\mu_2)} \\
 & + \frac{ \mu_0-\mu_2}{ \mu_1 (\nu_0-\lambda-\mu_1) (\nu_0-\lambda-\mu_2) (\nu_0-\lambda-\mu_0)} + \frac{ \mu_1-\mu_0}{ \mu_0 (\nu_0-\lambda-\mu_0)^2 (\nu_0-\lambda-\mu_1)} \\
 & + \frac{ \mu_0-\mu_1}{ \mu_2(\nu_0-\lambda-\mu_1)(\nu_0-\lambda-\mu_2)(\nu_0-\lambda-\mu_0)}  - \frac{(\mu_1-\mu_0)(\mu_2-\mu_0)}{ (\lambda-\nu_0)(\nu_0-\lambda-\mu_0)^2(\nu_0-\lambda-\mu_1)(\nu_0-\lambda-\mu_2)}  \bigg]  \bigg\}.
\end{align*}

\subsection{Marginalized variance and covariance derivation}
\normalsize
Because the initial state is uncertain, the variances and covariances of $X_3, X_4$ can now be computed by marginalizing over the initial barcoding state.
The marginalized means follow trivially by linearity and the law of total expectation: for instance,
$$ \E [X_3(t)] = \pi \E \left[ X_3(t) | \mb{X}(0) = (1,0,0,0) \right] + (1 - \pi) \E \left[ X_3(t) | \mb{X}(0) = (0,1,0,0) \right] = \pi M_{3|1} + (1-\pi) M_{3|2}. $$
 Dropping the dependence on $t$ for notational simplicity, we use the law of total variance and law of total covariance to obtain the marginalized variance expressions
\begin{align}
 \text{Cov}(X_3, X_4) &= \pi^2 ( U_{34} - M_{3|1} M_{4 | 1} ) + (1 - \pi)^2 ( V_{34} - M_{3|2} M_{4|2} ) \nonumber \\
& + \pi(1-\pi) ( U_{34|1} + U_{34|2} - M_{3|2}M_{4 | 1} - M_{3|1}M_{4|2} ) \nonumber \\
\text{Var}(X_3) &= \pi(U_{33|1} + M_{3|1} ) + (1-\pi) (U_{33|2} + M_{3|2}) \nonumber \\
& - \pi^2 M_{3|1}^2 - (1-\pi)^2 M_{3|2}^2 - 2\pi(1-\pi) M_{3|1}M_{3|2} \nonumber\\
\text{Var}(X_4) &= \pi(U_{44|1} + M_{4 | 1} ) + (1-\pi) (U_{44|2} + M_{4|2}) \nonumber\\
& - \pi^2 M_{4 | 1}^2 - (1-\pi)^2 M_{4|2}^2 - 2\pi(1-\pi) M_{4 | 1}M_{4|2} \label{eq:margvar}.
\end{align}

We now include the details behind Equation \eqref{eq:margvar} and derive the expressions in the general case with $K$ progenitors. Applying the law of iterated variance, the total variance for a type $i$ mature cell population is given by
\[ \V [X_i(t)] = \underbrace{\E [ \V [ X_i(t) | \mb{X}(0) ] ]}_{\text{(1)}} + \underbrace{\V[ \E [ X_i(t) | \mb{X}(0) ] ] }_{\text{(2)}}. \]
We drop the dependence on $t$ in intermediate steps for simplicity, and adopt the notation 
$$\E (X_{i|1}^2) =\E \left[ X_i^2 | \mb{X}(0) = (1,0,0,\ldots,0) \right], $$ 
and similarly use $\E [X_{i|j}]$ for expectations of $X_i(t)$ conditional on beginning with one initial type $j$ particle at $t=0$. 
With these conventions, the outer expectation over initial barcoding probability (1) simplifies to
\begin{align*}
\E [ \V [ X_i | \mb{X}(0) ] ] &= \E \left\{ \E [ X^2_i | \mb{X}(0) ]  - [\E [ X_i | \mb{X}(0)]]^2 \right\}\\
&=
 \pi_1 \E(X_{i|1}^2) + \ldots + \pi_{K} \E(X_{i|K}^2) - \pi_1 \left[\E(X_{i|1})\right]^2 + \ldots + \pi_{K} \left[\E(X_{i|K})\right]^2  \\
&= \sum_{k=1}^K \pi_k \E \left( X_{i|k} ^2 \right) - \sum_{k=1}^K \pi_k \left[\E\left( X_{i|k} \right) \right]^2.  
\end{align*}
Next, it is straightforward to expand (2) as 
\begin{align*}
\V[ \E [ X_i | \mb{X}(0) ] ] &= 
\E\{\E [ X_i | \mb{X}(0) ]\}^2 - \left(\E\{\E [ X_i | \mb{X}(0) ]\}\right)^2\\
&= \sum_{k=1}^K \pi_k  \left[\E\left(X_{i|k} \right)\right]^2  -  \left[\sum_{k=1}^K \pi_k \E(X_{i|k})\right]^2.
\end{align*}
Combining these simplifications (1) + (2), we arrive at the total variance expression marginalized over initial state:
\begin{equation}\label{eq:totalvar} \V [X_i(t)] = \sum_{k=1}^K \pi_k \E [ X_{i|k} ^2 ] - \sum_{k=1}^K \pi_k^2 \E\left[\left(X_{i|k}\right)\right]^2 - 2 \sum_{j > k} \pi_j \pi_k \E[ X_{i|j}] \E[ X_{i|k} ] . \end{equation}
Analogously to \eqref{eq:margvar} for the four-type model, this expression is directly related to the closed form solutions we obtain from solving the systems of moment differential equations. In terms of moment expressions, \eqref{eq:totalvar} becomes
\[ \V [X_i(t)] = \sum_{k=1}^K \pi_k [ U_{ii | k}(t) + M_{i | k}(t) ] -  \sum_{k=1}^K \pi_k^2 M_{i|k}(t)^2 - 2 \sum_{j > k} \pi_j \pi_k M_{i|k}(t) M_{i|j}(t). \]
The marginal covariance expressions are then obtained exactly analogously, applying the law of total covariance instead of the law of total variance. The covariances are given by
\begin{align*}
\cov{[X_i(t), X_j(t)]} &= \sum_{k=1}^K \pi_k \E[ X_{i|k} X_{j|k}] - \sum_{k=1}^K \pi_k^2 \E[X_{i|k}] \E [X_{j|k} ] - \sum_{k\neq l} \pi_k \pi_l \E[X_{i|k}] \E[X_{j|l}] \\ 
&= \sum_{k=1}^K \pi_k  U_{ij | k}(t)  -  \sum_{k=1}^K \pi_k^2 M_{i|k}(t) M_{j|k}(t) -  \sum_{k \neq l} \pi_k \pi_l M_{i|k}(t) M_{j|l}(t) .
\end{align*}
Given these marginalized variance and covariance expressions, incorporating the hypergeometric sampling distribution to obtain covariance and variance between read data $\mb{Y}$ is accomplished similarly. Applying the law of total (co)variance with respect to the sampling distribution yields
\begin{align*}
\cov{(Y_m,Y_n)} &= \E[\cov{(Y_m, Y_n)}| \mb{X}] + \cov{[ \E(Y_m | \mb{X}), \E(Y_n|\mb{X})]} \\
&= 0 + \cov{\left(\frac{b_m}{B_m} X_m, \frac{b_n}{B_n} X_n\right)} =\frac{b_m b_n}{B_m B_n} \cov{(X_m,X_n)} ,
\end{align*}
\begin{align*}
\V(Y_m) &= \E[ \V(Y_m | \mb{X})] + \V[ \E(Y_m | \mb{X})] \\
&= \E \left[ \frac{b_m X_m}{B_m} \frac{B_m - X_m}{B_m} \frac{B_m - b_m}{B_m - 1} \right] + \V \left(\frac{b_m X_m}{B_m} \right) \\
&= \frac{b_m (B_m - b_m)}{B_m^2 (B_m - 1)} \E(B_m X_m - X_m^2) + \frac{b_m^2}{B_m^2} \V(X_m).
\end{align*}

\paragraph{Treating $B_m$ as constant:} 
The term $B_m(t) = \sum_p X_m^p(t)$ is treated as a constant throughout above derivations. To motivate this approximation, we observe that its effect becomes negligible as the number of random variables $N$ appearing as summands grows. Given IID random variables $X_i$ with mean $\mu$,
\begin{equation}\label{eq:example} \text{Var} \left[ \frac{X_1}{\sum_{i=1}^N X_i} \right] = \text{Var} \left[ \frac{X_1/N}{1/N \sum_{i=1}^N X_i} \right] \approx \frac{1}{(N\mu)^2} \text{Var}(X_1), \end{equation}
where we have $\lim_{N \rightarrow \infty} \sum X_i = N\mu$, $\lim_{N \rightarrow \infty} \text{Var}(\sum X_i/N ) = 0$ by the of large numbers and the central limit theorem. 
This informally illustrates that treating the sum $\sum_{i=1}^N X_i$ as constant becomes more appropriate as $N$ grows. As we sum over almost $10,000$ IID barcodes, this regime is well-justified.  
Moreover, we have empirical evidence that treating $B_m$ as constant throughout yields good estimates on simulated data in Section \ref{sec:sim}. That is, the data were are simulated unconditional on $B_m$, and plugging in the observed values of $B_m$ at each observation time leads to accurate parameter estimates. 
To isolate this effect and illustrate it more explicitly, we include an additional simple simulation study. Given a true mean vector $\boldsymbol\mu$ and covariance matrix 
$$\Sigma = \begin{bmatrix}
    2      & 3  \\
    3      & 6  \\ \end{bmatrix}= \begin{bmatrix}
    a      & b  \\
    b      & c  \\ \end{bmatrix},$$ 
we generate IID bivariate normal random variables $(X_1,Y_1) \ldots, (X_N,Y_N)$. We compute the sample covariance $\hat{\text{Cov}}\left( \frac{X}{\sum_{i=1}^N X_i}, \frac{Y}{\sum_{i=1}^N Y_i}\right)$, and appropriately scale the entries following the form of \eqref{eq:example}, producing the empirical estimate 
 $$\hat{\Sigma} = \begin{bmatrix}
    \hat{a}      & \hat{b}  \\
    \hat{b}      & \hat{c}  \\ \end{bmatrix} = \hat{\text{Cov}}\left( \frac{X}{\sum_{i=1}^N X_i}, \frac{Y}{\sum_{i=1}^N Y_i}\right) \ast \begin{bmatrix}
    \frac{1}{N^2\mu_1^2}      & \frac{1}{N^2\mu_1 \mu_2}  \\
    \frac{1}{N^2\mu_1 \mu_2}      & \frac{1}{N^2\mu_2^2}  \\ \end{bmatrix}    
    .$$
Here $\ast$ denotes entry-wise multiplication. The table below displays the estimates of the entries of $\hat{\Sigma}$ as we repeat this procedure over a range of values of $N$, demonstrating that the information provided by the sum $\sum_{i=1}^N X_i$ has an increasingly negligible effect when $N$ is large. 
\begin{table}[ht]
\centering
\begin{tabular}{rrrrrrrrr}
  \hline
$N$ & 5 & 10 & 20 & 50 & 100 & 1,000 & 10,000 & True\\ 
  \hline
$\hat{a}$ & 1.66 & 1.78 & 1.85 & 1.95 & 1.98 & 1.97 & 1.99 & 2\\ 
$\hat{b}$ & 2.47 & 2.67 & 2.79 & 2.90 & 2.96 & 3.01 & 2.99 & 3\\
$\hat{c}$ & 4.85 & 5.37 & 5.62 & 5.75 & 5.92 & 6.06 & 5.98 & 6\\ 
   \hline
\end{tabular}
\caption{Estimates are averaged over $10,000$ trials.}
\end{table}

\section{Unconstrained parametrization of initial barcoding vector:} For models with multiple progenitor types, the initial barcoding probabilities are represented as a vector $\boldsymbol\pi = (\pi_1, \ldots, \pi_k)$ where $\pi_1$ denotes the probability of starting as an HSC, and $\pi_i$ denotes the probability of starting as a type $i$ progenitor for $i = 2, \ldots, K $. These parameters $\pi_i$ are naturally constrained to a probability simplex, but in practice we reparameterize by borrowing from a technique used in multinomial logistic regression by defining a set of variables $\gamma_i := \ln \left(\pi_i / \pi_K\right)$ for $i = 1, \ldots, K-1$. Then notice $\pi_i = \pi_K e^{\gamma_i}$ for all $i \leq K-1$, and letting $\pi_K = \frac{1}{1 + \sum_{i=1}^{K-1} e^{\gamma_i} }$, we ensure the simplex constraint that $\sum_{i =1}^K \pi_i = 1$. This enables us to equivalently consider the vector $\boldsymbol\gamma = (\gamma_1, \ldots,\gamma_{K-1})$ as parameters instead of $\boldsymbol\pi$, and because $\gamma_i$ vary freely in $\mathbb{R}$, we no longer need to add a constraint to the optimization problem.

\section{Detailed simulation results}
Here, we include detailed tables of true parameters used to initiate simulation as well as median estimates, median absolute deviations, and standard deviations corresponding to the simulation study design discussed in section \ref{sec:sim} for all model structures depicted in Figure \ref{fig:allmodels}. Some models depicted in Figure \ref{fig:allmodels} are identical in simulation study --- for instance, models (c) and (d) have no difference when final types are arbitrary. We also note that estimates reported in Table \ref{tab:results_5_2} correspond to the results plotted in Figure \ref{fig:synthEst} in the main text.

\begin{table}[ht]
\centering
\begin{tabular}{rrrrrrrr}
  \hline
 & $\lambda$ & $\nu_a$ & $\mu_a$ & $\nu_1$ & $\nu_2$ & $\nu_3$ & $\pi_a$ \\ 
  \hline
True & 0.0280 & 0.0200 & 0.0080 & 36 & 15 & 7 & 0.9000 \\ 
Median & 0.0283 & 0.0194 & 0.0086 & 34.84 & 14.18 & 6.624 & 0.8959 \\ 
MAD & 0.0008 & 0.0009 & 0.0021 & 6.31 & 2.797 & 1.167 & 0.0201 \\ 
SD & 0.0008 & 0.0010 & 0.0021 & 10.33 & 4.623 & 1.993 & 0.0199 \\ 
   \hline
\end{tabular}
\caption{Results of estimation on synthetic data from a model with three mature types and one common progenitor compartment, i.e. Model (a) in Figure \ref{fig:allmodels} of the main text, in terms of medians, standard deviations (SD), and median absolute deviations (MAD). With fixed death rates at $\mu_1=0.24, \mu_2=0.14, \mu_3=0.09$, estimates are very close to true parameters used to simulate the data. Recall $\pi_a$ denotes the proportion barcoded as progenitors, while $\pi_1 = 1-\pi_a$ is the proportion marked at the HSC stage. }
\end{table}

\begin{table}[H]
\begin{center}
\begin{tabular}{lrrrrrrrrr} 
 & $\lambda$ & $\nu_a$ & $\mu_a$ & $\nu_1$ & $\nu_2$ & $\nu_3$ & $\nu_4$ & $\nu_5$ & $\pi_a$ \\ 
 \hline
True & 0.0285  & 0.0200 & 0.0080 & 36.00 & 15.00 & 10.00 & 20.00 & 7.000 & 0.9000  \\ 
Median  &  0.0284 & 0.0200 & 0.0076 & 37.16 & 15.54 & 10.35 & 20.69 & 7.246 & 0.9021 \\
MAD & 0.0007 & 0.0011 & 0.0016 & 5.851 & 2.568 & 1.693 & 3.399 & 1.178 & 0.0153 \\
SD & 0.0025 & 0.0019 & 0.2800 & 11.84 &  3.568 & 2.504 &  4.574 &  1.994 &  0.0465 \\ 
\hline
\end{tabular}
\end{center}
\caption{Model with five mature types and one common progenitor compartment, i.e. Model (b) in Figure \ref{fig:allmodels}. Death rates fixed at $\mu_1=0.26, \mu_2=0.13, \mu_3=0.11, \mu_4=0.16, \mu_5=0.09$. }
\end{table}

\begin{table}[ht]
\centering \small
\begin{tabular}{lrrrrrrrrrrrr}
 & $\lambda$ & $\nu_a$ & $\nu_b$ & $\mu_a$ & $\mu_b$ & $\nu_1$ & $\nu_2$ & $\nu_3$ & $\nu_4$ & $\nu_5$ & $\pi_a$ & $\pi_b$ \\ 
  \hline
 True & 0.0285 & 0.0130 & 0.0070 & 0.0050 & 0.0040 & 36  & 15  & 10  & 20 & 7 & 0.60 & 0.30 \\
Med. & 0.0286 & 0.0130 & 0.0069 & 0.0045 & 0.0043 & 38.01 & 16.29 & 10.92 & 19.64 & 6.65 & 0.6333 & 0.2706 \\ 
  MAD & 0.0005 & 0.0008 & 0.0006 & 0.0021 & 0.0013 & 13.35 & 5.826 & 3.894 & 2.240 & 1.277 & 0.1399 & 0.1194 \\ 
  SD & 0.0006 & 0.0007 & 0.0007 & 0.0019 & 0.0012 & 17.61 & 7.828 & 5.241 & 5.347 & 1.925 & 0.1388 & 0.1255 \\ 
\hline
\end{tabular}
\caption{Model with five mature types and two distinct progenitor compartments, i.e. Model (c) in Figure \ref{fig:allmodels}. In this model, progenitor $a$ gives rise to type $1$ and $2$ mature cells, and $b$ produces type $3,4, $ and $5$ type cells. Estimates remain accurate in this parameter rich setting with multiple progenitor compartments. These correspond to estimates plotted in Figure~\ref{fig:synthEst} in the main text.}
\label{tab:results_5_2}
\end{table}

\begin{table}[ht]
\centering
\begin{tabular}{rrrrrrrrrrrrrrrr}
  \hline
 & $\lambda$ & $\nu_a$ & $\nu_b$ & $\nu_c$ & $\mu_a$ & $\mu_b$ & $\mu_c$ & $\nu_1$ \\ 
  \hline
True & 0.0500 & 0.0280 & 0.0140 & 0.0070 & 0.0080 & 0.0060 & 0.0020 & 40.0000 \\ 
Median & 0.0539 & 0.0303 & 0.0150 & 0.0075 & 0.0091 & 0.0058 & 0.0034 & 40.7977 \\ 
MAD & 0.0081 & 0.0047 & 0.0032 & 0.0016 & 0.0038 & 0.0072 & 0.0041 & 11.8020 \\ 
SD & 0.0143 & 0.0080 & 0.0052 & 0.0024 & 0.0037 & 0.0060 & 0.0053 & 18.0492 \\ 
   \hline
 & $\nu_2$ & $\nu_3$ & $\nu_4$ & $\nu_5$ & $\pi_a$ & $\pi_b$ & $\pi_c$ & \\
  \hline
True & 18.0000 & 14.0000 & 20.0000 & 8.0000 & 0.5500 & 0.2000 & 0.1500 \\ 
Median & 18.1527 & 17.7127 & 26.4716 & 10.6550 & 0.5595 & 0.2017 & 0.1578 \\ 
MAD & 5.0599 & 7.8044 & 9.4919 & 5.7547 & 0.0412 & 0.0120 & 0.0106 \\ 
SD & 7.0998 & 6.6657 & 8.8583 & 157.5674 & 0.0369 & 0.0159 & 0.0137 \\
   \hline
\end{tabular}
\caption{Synthetic data from a model with five mature types and three oligopotent and unipotent progenitors, i.e. Model (f) in Figure \ref{fig:allmodels}. Death rates fixed at $\boldsymbol\mu = (0.24, 0.13, 0.12, 0.18, 0.1)$. While the standard deviation reveals influence of extreme outliers on the estimate or $\nu_4$, median estimates are again accurate in a parameter rich model, and reasonably stable in terms of MAD.}
\end{table}

\subsection{Model misspecification experiments}
Tables \ref{tab:mis_5_3} and \ref{tab:mis_5_1} display the estimates obtained under over specified and misspecified models, along with objective values of the loss function at converged estimates; these correspond to total $\ell_2$ loss between fitted and observed correlations. Note that these estimates correspond to the correlation plots displayed in Figure \ref{fig:misPlots} in the main text.
\begin{table}[H]
\centering
\begin{tabular}{rrrrrrrrr}
 & $\lambda$ & $\nu_a$ & $\nu_b$ & $\nu_c$ & $\mu_a$ & $\mu_b$ & $\mu_c$ & $\nu_1$ \\ 
  \hline
Med. & 0.19365 & 0.05938 & 0.05475 & 0.00002 & 0.01136 & 0.19085 & 0.00056 & 56.89686 \\ 
  MAD & 0.06633 & 0.02942 & 0.02433 & 0.00003 & 0.01683 & 0.28294 & 0.00083 & 38.64162 \\ 
  SD & 0.07195 & 0.03583 & 0.03360 & 0.00015 & 0.59207 & 0.95299 & 0.00226 & 238.53322 \\ 
   & $\nu_2$ & $\nu_3$ & $\nu_4$ & $\nu_5$ & $\pi_a$ & $\pi_b$ & $\pi_c$ & Objective \\ 
  \hline
Med. & 22.10742 & 16.70475 & 33.70443 & 11.65951 & 0.00121 & 0.00038 & 0.83830 & 2.90319 \\ 
  MAD & 14.61860 & 9.40493 & 19.63489 & 5.62197 & 0.00179 & 0.00057 & 0.07087 & 0.75504 \\ 
  SD & 11.94989 & 7.67183 & 15.40404 & 5.04462 & 0.01796 & 0.00948 & 0.08250 & 1.08521 \\ 
   \hline
\end{tabular}
\caption{Model fit in overspecified case with three progenitors: note that the objective value is higher than the correct specification, and note that the estimates seem more spread apart than the correctly specified inference while representative of the overall shape of true correlation profiles. }
\label{tab:mis_5_3}
\end{table}

\begin{table}[H]
\centering
\begin{tabular}{rrrrrrrrrrr}
 & $\lambda$ & $\nu_a$ & $\mu_a$ & $\nu_1$ & $\nu_2$ & $\nu_3$ & $\nu_4$ & $\nu_5$ & $\pi_a$ & Objective \\ 
  \hline
Med. & 0.131 & 0.00468 & 0.0332 & 71.1 & 30.0 & 20.6 & 0.000 & 0.000 & 1.000 & 21.358 \\ 
  MAD & 0.0091 & 0.0041 & 0.0123 & 29.9 & 12.7 & 7.87 & 0.00000 & 0.00000 & 0.00001 & 0.221 \\ 
  SD & 0.0096 & 0.0078 & 0.0149 & 23.7 & 9.91 & 6.43 & 0.00000 & 0.00000 & 0.00001 & 0.227 \\ 
   \hline
\end{tabular}
\caption{Underspecified model fit. Interestingly, this model seems to correctly identify that types $1,2,3$ are linked from a common progenitor, but because one shared progenitor is not compatible with the observed correlations, and in particular cannot explain negative correlations between types from distinct lineages, the model assigns almost zero mass to rates $\nu_4, \nu_5$ of producing the other mature types. The solution seems to be strongly a boundary solution with all barcoded cells starting in the progenitor compartment, resulting in a very poor objective function value.}
\label{tab:mis_5_1}
\end{table}

\begin{table}[H]
\centering \small
\begin{tabular}{rrrrrrrrrrrr}
 & $\lambda$ & $\nu_a$ & $\nu_b$ & $\mu_a$ & $\mu_b$ & $\nu_1$ & $\nu_2$ & $\nu_3$ & $\pi_2$ & $\pi_3$ & Obj. \\ 
  \hline
Med. & 0.0286 & 0.0130 & 0.0080 & 0.0077 & 0.0014 & 31.43 & 21.32 & 46.57 & 0.533 & 0.358 & $9.093 \times 10^{-5}$ \\ 
  MAD & 0.0007 & 0.0009 & 0.0011 & 0.0022 & 0.0020 & 6.595 & 5.097 & 29.47 & 0.1096 & 0.1009 & $4.629 \times 10^{-5}$ \\ 
  SD & 0.0067 & 0.0043 & 0.0028 & 0.0026 & 0.0022 & 22.31 & 21.54 & 63.07 & 0.1791 & 0.1943 & $6.950 \times 10^{-5}$ \\ 
   \hline
\end{tabular}
\caption{Results corresponding to three grouped mature cell compartments with correctly specified progenitor structure. Note the objective value here is orders of magnitude lower than the five-type models with misspecified progenitor structures, suggesting that lumping mature types is a justifiable model simplification compared to the tradeoff of specifying a richer model with flawed assumptions on the intermediate structure. }
\label{tab:mis_3_2}
\end{table}

\subsection{Tables of complete estimated parameters fitted to lineage barcoding data}
\begin{table}[H]
\begin{center}
\begin{tabular}{lllllll}
 \hline
Par & (a) & (b) & (c) & (d)& (e) & (f) \\
\hline
$\hat\lambda$ & 0.0593  & 0.0867 & 0.4360 & 0.3644 & 0.2271 & 0.3198 \\ 
$\hat\nu_a$ 	& $1.00 \mathrm{e}$-6 & $1.80 \mathrm{e}$-7 & 0.4090 & 0.3521 & 0.1446 & 0.0033 \\
$\hat\nu_b$ &		&	& 0.0257 & 0.0121 & 0.0725 & 0.3131 \\
$\hat\nu_c$ &		&	&  & & 0.0101 & 0.0033 \\
$\hat\mu_a$ & $7.95 \mathrm{e}$-6 & $0.0367$ & 1.150 & 4.096 & 0.7037 & 0.1449\\
$\hat\mu_b$ &		&	& 4.023 & 3.699 & 4.022 & $1.253 \mathrm{e}$-3\\
$\hat\mu_c$ &		&	& & & 3.602 & 1.434\\
$\hat\nu_1$ & 2042.0  & 1486.3 & 1305.5 & 866.1 & 1896.8 & 1959.09 \\
$\hat\nu_2$ & 434.7 & 1764.3 & 201.4 & 391.3 & 221.3 & 560.4\\
$\hat\nu_3$ & 147.4  & 74.0 & 113.6 & 264.4 & 112.3 & 127.5\\
$\hat\nu_4$ &   & 326.4 & 448.7 & 299.5 & 417.1 & 287.9\\
$\hat\nu_5$ &   & 17.9 & 17.0 & 54.1 & 79.3 & 104.2 \\
$\hat\pi_a$ & 0.861  & $0.87$ & 0.870 & 0.870 & 0.0 & 0.0\\
$\hat\pi_b$ & & &  0.0 & 0.0 & 0.0 & 0.870\\
$\hat\pi_c$ & & &  & & 0.870 & 0.0 \\
\hline
Loss & 0.4071 & 1.653 & 3.465 & 3.330 & 3.836 & 2.91 \\
\hline
\end{tabular}
\end{center}
\caption{Parameter estimates for all models displayed in Figure \ref{fig:allmodels}. Model (a) has fixed deaths $(0.6, 0.04, 0.4)$. All other models have fixed death rates $(0.8, 0.3, 0.04, 0.08, 0.4)$. Recall sum of progenitor barcoding proportions fixed to be $0.87$ for models (b)-(f). } 
\label{tab:realdata}
\end{table}

\begin{table}[H]
\begin{center}
\footnotesize
\begin{tabular}{lllllll}
 \hline
Par & (a) & (b) & (c) & (d)& (e) & (f) \\
\hline
$\hat\lambda$ & $(0.003,0.109)$  & $(0.077,0.163)$  & $(0.196,1.168)$  & $(0.207,0.640)$  & $(0.132,0.752)$ & $(0.174,0.449)$ \\ 
$\hat\nu_a$ 	& $(0.0,0.004)$ & $(0.0,0.001)$  & ($0.085,1.148)$ & $(0.131,0.611)$ & $(0.014,0.617)$ & $(0.074,0.396)$\\
$\hat\nu_b$ &  &  & $(0.006,0.168)$ & $(0.006,0.112)$ & $(0.015,0.486)$ & $(0.021,0.154)$ \\
$\hat\nu_c$ &  &  &  &  & $(0.000,0.019)$ & $(0.000,0.010)$\\
$\hat\mu_a$ & $(0.0, 0.002)$ & $(0.028,0.046)$ & $(0.000,3.879)$  & $(0.267,3.603)$  & $(0.0,2.854)$ & $(0.246,2.651)$\\
$\hat\mu_b$ &  &  & $(0.434,4.022)$ &  $(0.437,4.102)$ & $(0.447,4.023)$ & $(0.811,4.543)$\\ 
$\hat\mu_c$ &  &  &  &  & $(0.102,4.103)$ & $(0.283,4.100)$\\
$\hat\nu_1$ & $(956.0,2239.9)$ & $(830.1,1838.6)$ & $(627.7,1482.2)$ & $(613.3,1487.3)$ & $(600.9,1495.0)$ & $(615.7,1474.2)$ \\
$\hat\nu_2$ & $(52.0,488.7)$ & $(1021.9,2055.3)$  &  $(131.8,521.4)$ & $(135.6,344.6)$ & $(187.9,477.2)$ & $(255.9,448.7)$ \\
$\hat\nu_3$ & $(39.7,148.2)$ & $(60.7,99.8)$ & $(30.6,294.6)$ & $(134.5,305.0)$  & $(4.279,291.3)$  & $(126.3,297.8)$\\
$\hat\nu_4$ &  & $(275.2,470.7)$ & $(146.2,558.7)$ & $(126.4,297.4)$ & $(127.8,321.7)$  & $(128.2,295.7)$\\
$\hat\nu_5$ &  & $(10.1,44.65)$ & $(3.786,9.559)$ & $(1.137,10.63)$ & ($6.488,84.9)$  & $(26.4,74.0)$ \\
$\hat\pi_a$ & $(0.017,0.861)$ & $(0.87,0.87)$ & $(0.0,.0.599)$ & $(0.0,.598)$ & $(0.0,0.038)$  & $(0.000,0.001)$\\
$\hat\pi_b$ &  &  & $(0.0,0.999)$ & $(0.0,0.999)$ & $(0.0,1.0)$ & $(0.999,1.0)$ \\
$\hat\pi_c$ &  &  &  &  & $(0.0,1.0)$  & $(0.000,0.000)$\\
\hline
Loss & $(0.352,0.696)$ & $(1.485,2.341)$ & $(2.591,4.771)$ & $(2.472,4.489)$ & $(3.092,5.763)$ & $(2.566,4.777)$ \\
\hline
\end{tabular}
\end{center}
\caption{Corresponding $95\%$ confidence intervals produced via nonparametric bootstrap of $2500$ replicate datasets. Recall sum of progenitor barcoding proportions fixed to be $0.87$ for models (b)-(f).} 
\label{tab:realdataCI}
\end{table}

\subsection{Additional fitted correlation profiles fitted to lineage barcoding dataset}
\begin{figure}
\centering
\includegraphics[width = 0.85\textwidth]{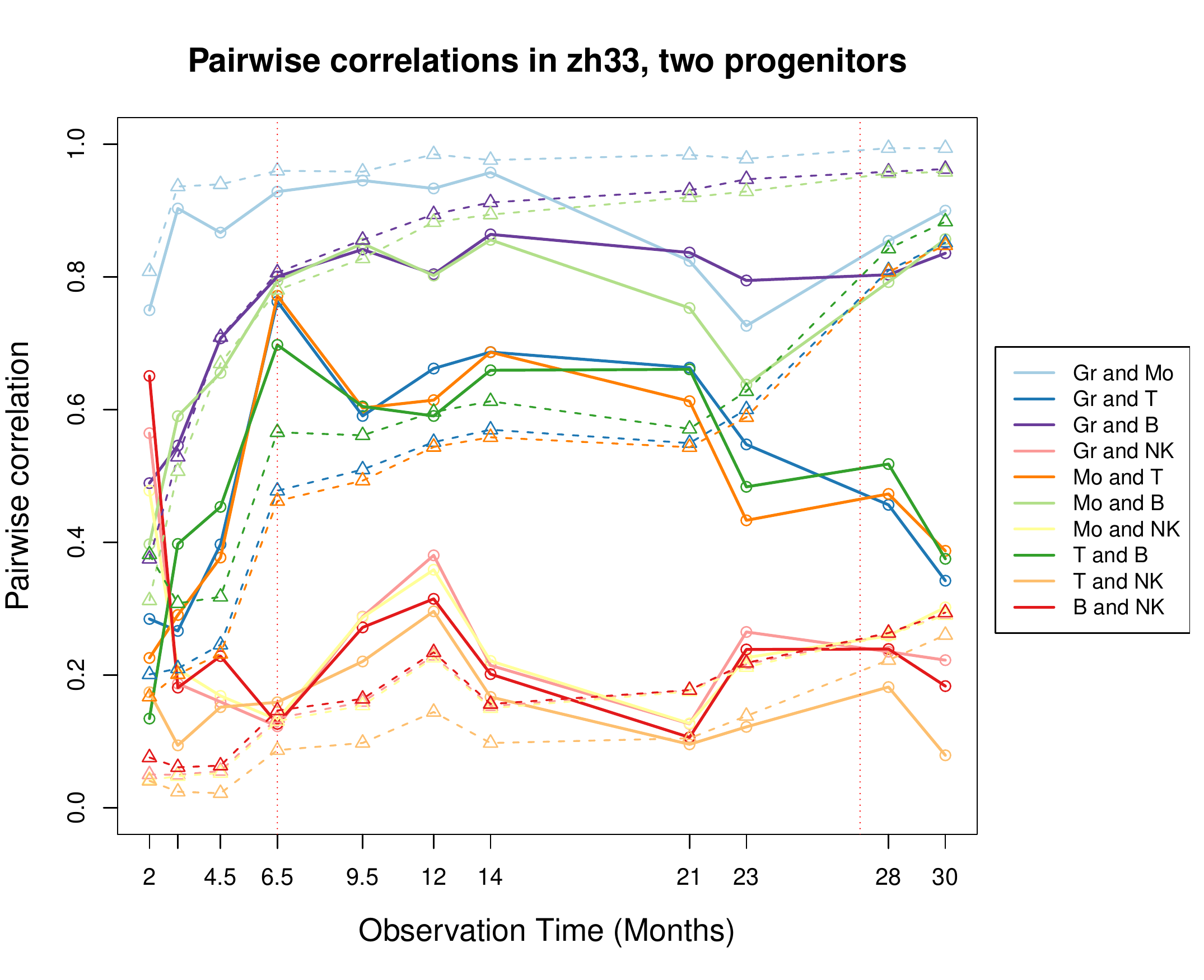} 
\caption[Fitted correlation curves in two-progenitor model to rhesus macaque data]{Fitted curves for real data to model with two progenitors, corresponding to model (c) displayed in Figure \ref{fig:allmodels}. The ``misgrouped" fitted curves apparent after 23 months visually suggest the misspecification in designating specialized oligopotent progenitors.} \label{fig:realdata2}
\end{figure}

\begin{figure}
\centering
\includegraphics[width = 0.85\textwidth]{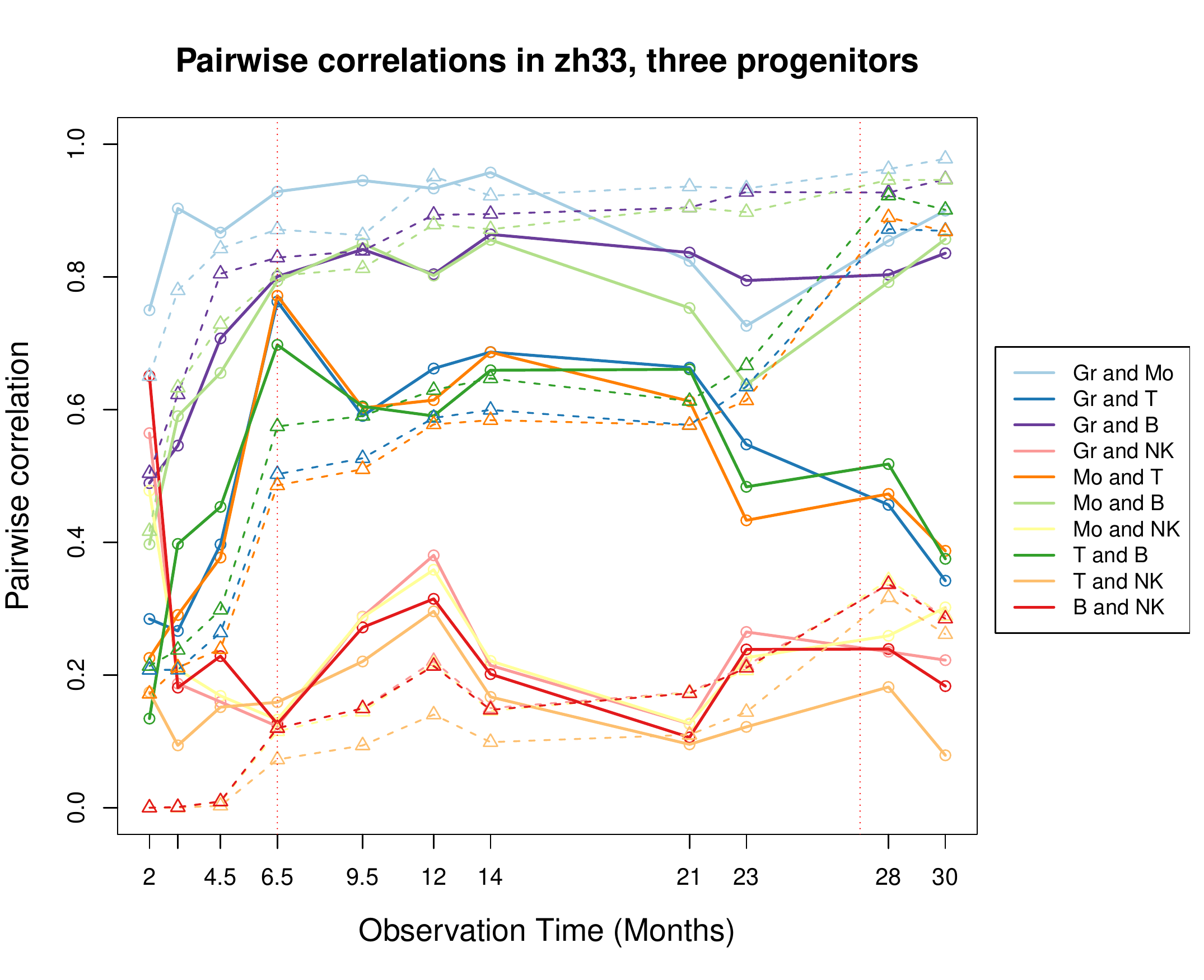}
\caption[Fitted correlation curves in three-progenitor model to rhesus macaque data]{Fitted curves for real data in model with three specialized progenitors, i.e. model (e) in Figure \ref{fig:allmodels}. Again, a misgrouping is visually apparent in fitted curves after 23 months} \label{fig:realdata3}
\end{figure}

\newpage

\end{document}